\documentclass[aps,prd,10pt,twocolumn,floatfix,
  noshowpacs,preprintnumbers,superscriptaddress
  ,nofootinbib
  ]{revtex4-1}
\usepackage[utf8]{inputenc}
\usepackage{amsmath}
\usepackage{amssymb}
\usepackage[T1]{fontenc}
\usepackage{bbold}

\usepackage{enumerate}
\usepackage{amsmath}
\usepackage{tikz}
\usepackage[compat=1.1.0]{tikz-feynman}
\usepackage{feynmf}
\usepackage{slashed}
\usepackage{braket}
\usepackage{url}
\usepackage{multirow}
\usepackage{natbib}
\usepackage{graphicx}
\usepackage[pdftex,bookmarks,linktocpage,pdfpagelabels,plainpages=false,hyperfigures,linkcolor=blue,citecolor=blue]{hyperref} %for hypertext links
\hypersetup{colorlinks=true}

\usepackage{mathrsfs,amssymb}  
\usepackage{cancel}
\usepackage[normalem]{ulem}

\begin{document}
\bibliographystyle{apsrev4-1}
\title{Solutions to the MiniBooNE Anomaly from New Physics in Charged Meson Decays}

\author{Bhaskar Dutta}
\affiliation{Mitchell Institute for Fundamental Physics and Astronomy, Department of Physics and Astronomy, Texas A\&M University, College Station, TX 77845, USA}

\author{Doojin Kim}
\affiliation{Mitchell Institute for Fundamental Physics and Astronomy, Department of Physics and Astronomy, Texas A\&M University, College Station, TX 77845, USA}

\author{Adrian Thompson}
\affiliation{Mitchell Institute for Fundamental Physics and Astronomy, Department of Physics and Astronomy, Texas A\&M University, College Station, TX 77845, USA}

\author{Remington T. Thornton}
\affiliation{Los~Alamos~National~Laboratory,~Los~Alamos,~NM~87545,~USA}

\author{Richard G.~Van~de~Water}
\affiliation{Los~Alamos~National~Laboratory,~Los~Alamos,~NM~87545,~USA}

\begin{abstract}
    We point out that production of new bosons by charged meson decays can greatly enhance the sensitivity of beam-focused accelerator-based experiments to new physics signals. This enhancement arises since the charged mesons are focused and their three-body decays do not suffer from helicity suppression in the same way as their usual two-body decays. 
    As a realistic application, we attempt to explain the MiniBooNE low energy excess utilizing this overlooked mechanism, uniquely realizing dark-sector interpretations as plausible solutions to the excess. 
    As proof of the principle, we consider two well-motivated classes of dark-sector models, models of vector-portal dark matter and models of long-lived (pseudo)scalar. We argue that the model parameter values to accommodate the excess are consistent with existing limits and that they can be tested at current and future accelerator-based neutrino experiments.
\end{abstract}

\preprint{MI-HET-766}
\preprint{LA-UR-21-30532}
\maketitle

\noindent {\bf Introduction}.
The MiniBooNE excess of electron-like events at $4.8\sigma$~\cite{MiniBooNE:2008yuf,MiniBooNE:2018esg,MiniBooNE:2020pnu} has been considered as one of the renowned phenomena indicative of the existence of new physics beyond the Standard Model (SM). 
Although a recent study claims that a careful estimate of systematics associated with major backgrounds would reduce the confidence level~\cite{Brdar:2021cgb}, the excess remains evident and requires a reasonable explanation.
Furthermore, the recent MicroBooNE result~\cite{MicroBooNE:2021zai} constrains the $\Delta \to N\gamma$ background more stringently, disfavoring the possibility of its $\sim\!3$ times more enhanced branching ratio (BR)~\cite{MiniBooNE:2020pnu}. This observation advocates the need for new physics to explain the MiniBooNE $e$/$\gamma$-like excess and the MicroBooNE suggestive coherent-like scattering excess, both at low energies relative to the beam energy. 
Various new physics scenarios have been proposed to explain the anomaly. Amongst them, neutrino-based solutions~\cite{Sorel:2003hf,Karagiorgi:2009nb,Collin:2016aqd,Giunti:2011gz,Giunti:2011cp,Gariazzo:2017fdh,Boser:2019rta,Kopp:2011qd,Kopp:2013vaa,Dentler:2018sju,Abazajian:2012ys,Conrad:2012qt,Diaz:2019fwt,Asaadi:2017bhx,Karagiorgi:2012kw,Pas:2005rb,Doring:2018cob,Kostelecky:2003cr,Katori:2006mz,Diaz:2010ft,Diaz:2011ia,Gninenko:2009ks,Gninenko:2009yf,Bai:2015ztj,Moss:2017pur,Bertuzzo:2018itn,Ballett:2018ynz,Fischer:2019fbw,Moulai:2019gpi,Dentler:2019dhz,deGouvea:2019qre,Datta:2020auq,Dutta:2020scq,Abdallah:2020biq,Abdullahi:2020nyr,Liao:2016reh,Carena:2017qhd,Abdallah:2020vgg} 
have received particular attention, as they also accommodate the observation that the MiniBooNE off-target mode does not show any appreciable excess~\cite{MiniBooNEDM:2018cxm}.
This particularly challenges dark-sector interpretations including the scenario in which dark-matter production occurs dominantly from the decay of neutral mesons (e.g., $\pi^0$) together with kinetic mixing between the SM photon and a dark-sector $U(1)$ gauge particle, as they would give rise to a corresponding excess in the off-target mode~\citep{MiniBooNEDM:2018cxm,Jordan:2018qiy}.  
In addition, most of the solutions, including the neutrino-based explanations, are potentially in conflict with null signal observations at other neutrino experiments including CHARM-II, MINER$\nu$A, and T2K~\cite{Arguelles:2018mtc,Brdar:2020tle}. 

In light of this situation, we point out that a hypothetical decay of charged mesons (e.g., $\pi^\pm$ and $K^\pm$) to a new mediator [vector/(pseudo)scalar] together with a $\ell\nu_\ell$ pair 
can provide a robust solution to the MiniBooNE anomaly that is nearly immune to the aforementioned issues, taking a few benchmark scenarios. 
In particular, we demonstrate that this class of explanations can render dark-sector interpretations (e.g., dark matter and long-lived mediators) plausible solutions without involving neutrino-sector physics. 
We also briefly discuss the implications of the recent result from MicroBooNE on these solutions.
We further point out that ongoing and upcoming 
accelerator-based neutrino experiments 
can test some of them, confirming the MiniBooNE excess. 

\medskip

\begin{figure*}[t]
    \centering
    \includegraphics[width=15.8cm]{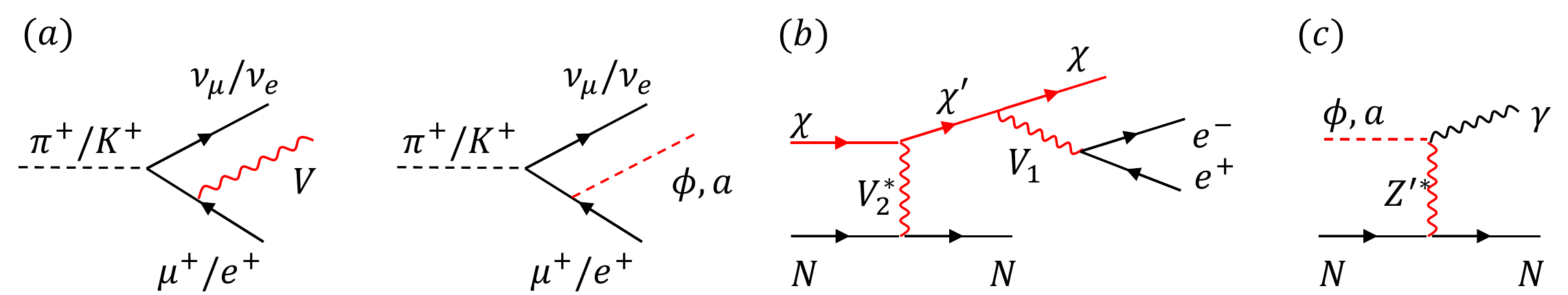}
    \caption{$(a)$: Three-body charged meson decay into a scalar, pseudoscalar, or vector. $(b)$: Dark-matter upscattering via a vector mediator. In the single-mediator case, $V_2=V_1$. $(c)$: ``Dark Primakoff'' scattering of a scalar/pseudoscalar $\phi$/$a$ via a 
    %heavy vector mediator light 
    $Z^\prime$.
    }
    \label{fig:3body}
\end{figure*}

\noindent {\bf New physics from the $\pi^\pm/K^\pm$ decays}.
The two-body decay process of charged pion or kaon, $\pi/K \to \ell \nu_\ell$ ($\ell=e,\mu$), is highly suppressed unlike the na\"{i}ve phase-space expectation, because the chiral nature of the decay products forces the angular momentum conservation to hold in limited phase space. However, once another decay product is added, the angular momentum conservation can be easily satisfied, allowing the process to fully exploit the decay phase space. 
As a consequence, a three-body decay involving (bosonic) mediator $\varphi$, $\pi/K \to \ell \nu_\ell \varphi$, shown in Fig.~\ref{fig:3body}$(a)$, can be sizable despite the additional phase-space suppression in the three-body process~\cite{Barger:2011mt,Carlson:2012pc,Laha:2013xua,Bakhti:2017jhm,Krnjaic:2019rsv} (see also Appendix~\ref{app:decay_width}).
We further find that this decay width enhancement of the three-body decay processes is so significant that the three-body processes can overcome the phase-space suppression and overwhelm the two-body processes with $\mathcal{O}(1)$ $\varphi$ coupling. 
In particular, if $\varphi$ is a massive vector, the corresponding enhancement becomes even more significant due to the existence of the longitudinal polarization mode~\cite{Carlson:2012pc,Laha:2013xua}. 
Of course, $\mathcal{O}(1)$ $\varphi$ coupling is unrealistic, as experimental measurements of the exotic decays of charged mesons set the upper limits of the BR of such decays to be $\sim\! 10^{-6}$-$10^{-9}$~\cite{ParticleDataGroup:2020ssz}.

However, one can satisfy this upper limit with realistic coupling values allowed by existing bounds for new mediators such as scalar, pseudo-scalar, and vector, and then expect a highly enhanced signal flux (either $\varphi$ itself or $\varphi$-induced) responsible for the MiniBooNE excess. 
Moreover, signal loss could be minimized since most of the charged $\pi/K$'s produced inside the MiniBooNE target are focused and directed to the MiniBooNE detector by the magnetic horn system before they decay.
Therefore, this class of explanations can be consistent with the null signal observation in the off-target mode, as in the neutrino-based interpretations. 
We will attempt to explain the MiniBooNE excess using the $\pi^\pm/K^\pm$ signal; for proof of the principle, we consider benchmark {\it simplified} models of vector and (pseudo)scalar mediators containing only the particle contents relevant to explaining the MiniBooNE anomaly. 
We emphasize that the larger mass gap between $K$ and the charged lepton opens more phase space than that in the $\pi$ decay so that the $K$-induced contribution can be comparable to or even larger than the $\pi$-induced despite its lower production rate, depending on the underlying model details. 

\medskip

\noindent {\bf Models}.
i) \underline{Vector-portal dark matter}: Using the three-body decay modes of $\pi^\pm/K^\pm$, one can consider inelastic dark-matter models.
For example, in a vector-portal dark-matter model, dark matter is produced by the decay of a dark-sector vector mediator (e.g., dark photon) and the scattering of the dark matter would contribute to the excess. The model is defined by the following interaction Lagrangian:
\begin{equation}
    \mathcal{L}_V \supset \hbox{$\sum_{i=1,2}$} \left( e \epsilon_i  J_{\rm EM}^\mu +g_i J_D^\mu + g^\prime_i J^{\prime\mu}_D\right)V_{i,\mu},
\end{equation}
where we generically assume the possibility that the vector mediator  responsible for production of dark matter (say, $V_1$) can differ from the mediator appearing in the scattering process of dark matter (say, $V_2$)~\cite{Dutta:2019nbn,Dutta:2020vop}. 
For illustration purposes, we assume that they are dark photons having different mixing parameters, $\epsilon_1$ and $\epsilon_2$. $J_{\rm EM}^\mu$ denotes the usual electromagnetic current in the SM, whereas $J_D^\mu$ and $J_D^{\prime\mu}$ describe the dark-sector currents involving only dark matter $\chi$ and involving dark matter and a heavier dark-sector state $\chi^\prime$, correspondingly. 
To be fully general, we here separately introduce dark-sector couplings, $g_{1,2}$ and $g^\prime_{1,2}$. Such models with detailed parameter-space analyses and UV completion were considered in the literature (e.g., Refs.~\cite{Tucker-Smith:2001myb,Izaguirre:2014dua,Giudice:2017zke,Dutta:2019fxn,Dutta:2020enk}).

Once dark photon $V_1$ is created by the charged meson decays in the MiniBooNE target, it promptly decays to a $\chi$ pair. For illustration purposes, we assume that ${\rm BR}(V_1 \to 2\chi)\!:\!\sum_f{\rm BR}(V_1 \to 2f)\!=\!50\%\!:\!50\%$ with $f$ running over all kinematically allowed SM fermions, and that mass relation $m_{V_1}<m_\chi+m_{\chi^\prime}$ holds so that $V_1\to \chi \chi^\prime$ is kinematically forbidden.  
A produced $\chi$ reaches the MiniBooNE detector and scatters off either a nucleon or a carbon nucleus into the heavier state $\chi^\prime$ via a $t$-channel exchange of $V_1$ (in the single-mediator scenario) or $V_2$ (in the double-mediator scenario). 
$\chi^\prime$ then decays back to a $\chi$ and an (on-shell) $V_1$ which subsequently decays into an electron pair in the fiducial volume of the detector. Here $V_1$ is significantly boosted, and therefore, the two Cherenkov rings induced by the electron pair typically overlap and appear single-ring-like within the detector angular resolution or as small mass in a two ring fit \footnote{Another route of obtaining a single electron-like signature is highly energy-asymmetric $e^+e^-$ pairs~\cite{Ballett:2018ynz} in the three-body decay of $\chi^\prime$ through an off-shell $V_1$.}.
We show this process in Fig.~\ref{fig:3body}$(b)$ and provide the scattering cross-section formulas including the nucleon and nucleus form factors in literature~(e.g., Refs.~\cite{Kim:2016zjx,Kim:2020ipj}).
See Appendix~\ref{app:dm_scat} for more details.

\smallskip

\noindent ii) \underline{Long-lived (pseudo)scalar}: 
In the next phenomenological model we consider a massive scalar $\phi$ or a massive pseudoscalar $a$ that both couple to muons, and a massive vector mediator $Z^\prime$, with couplings to quarks as well as to the (pseudo)scalar through the CP-(odd)even operators, respectively. Explicitly, the interaction Lagrangians for these mediators are
\begin{equation}
     \mathcal{L}_{S(P)} \supset g_n Z^\prime_\alpha \Bar{u} \gamma^\alpha u +
     \begin{cases}
     g_\mu \phi \Bar{\mu} \mu +
     \frac{\lambda}{4}\phi F^\prime_{\mu\nu} F^{\mu\nu} \\
     i g_\mu a \Bar{\mu} \gamma^5 \mu + \frac{\lambda}{4}aF^\prime_{\mu\nu} \Tilde{F}^{\mu\nu}
     \end{cases}
     + \textrm{h.c.},
     \label{eq:lagrangian_sp}
\end{equation}
where $g_\mu$, $g_n$, and $\lambda$ parametrize coupling strengths for the operators and where $F^\prime_{\mu\nu} \equiv \partial_\mu Z^\prime_\nu - \partial_\nu Z^\prime_\mu$ for $Z^\prime$.\footnote{These new bosons could appear, for example, in models with an extra gauged $U(1)_X$ symmetry and theories containing mixings between axion-like particles and dark photons (e.g., \cite{Kaneta:2016wvf,Kaneta:2017wfh,Hook:2021ous,deNiverville:2018hrc,Choi:2018mvk,Biswas:2019lcp,Kalashev:2018bra,Arias:2020tzl}).} 

While the coupling to muons facilitates the production of $\phi$ or $a$ from the charged mesons, similarly to the case of our vector portal model, in this situation we consider a long-lived $\phi$ or $a$ that scatters inside the fiducial volume of the MiniBooNE detector via a Primakoff-like process shown in Fig.~\ref{fig:3body}$(c)$. Like the Primakoff scattering of neutral-pion or axion-like particle, this is a coherent process, but it instead takes place via the $Z^\prime$'s coupling to quark matter. This allows for a nuclear coherence, which we have parameterized by the Helm nuclear form factor~\cite{Helm:1956zz}. 
The amplitudes for this process
are discussed in more detail in Appendix~\ref{app:dark_prim}.

\smallskip

In general, the long-lived (pseudo)scalar in Fig.~\ref{fig:3body}($c$) can be replaced by a long-lived vector mediator 
while a (pseudo)scalar replaces the $Z^\prime$, and the related phenomenology in the context of the MiniBooNE excess is qualitatively/quantitatively similar to that of Model ii).
Furthermore, depending on model details, one can envision the cases where the mediators in Fig.~\ref{fig:3body}($a$) can be attached to the $\pi^\pm/K^\pm/\nu$ legs as well as the $\ell^\pm$ leg.

\medskip

\noindent {\bf Simulation}.
Parameterizations of the charged meson fluxes at the MiniBooNE target are given in \cite{MiniBooNE:2008hfu} for the Sanford-Wang (SW) and Feynman-Scaling (FS) approaches to modeling the $\pi^\pm$ and $K^+$ fluxes, respectively. These parameterizations provide the double differential production cross-sections of $\pi^\pm$ and $K^+$ in the outgoing meson momentum $p$ and angle $\theta$ with respect to the beam axis, as a function of the incident proton momentum. 
We have validated the SW and FS parameterizations by reproducing the MiniBooNE-modeled neutrino fluxes~\cite{MiniBooNE:2008hfu} to within an $\mathcal{O}(1)$ normalization difference.
As the corresponding parametrization for $K^-$ is unavailable, we instead adapt the FS-based BMPT model~\cite{Bonesini:2001iz} to describe $K^-$ production inside the target. 

Most of the produced mesons enter the focusing-horn area where their momentum is (almost) aligned with the beam direction. The focusing-horn geometry allows mesons of $\theta \in (0.03, 0.21)$ radians to get focused~\cite{Schmitz:2008zz}. We therefore select the mesons produced within this angular range out of the mesons simulated according to the above-described parametrizations and assume that their momentum becomes fully parallel to the beam axis by the focusing horn. We then check whether the chosen meson decays within 50 meters before reaching the dump area, using the usual decay law. The kinematics of the $\pi^\pm/K^\pm$ three-body decay, which involves production of mediator $\varphi(= V_1, a, \hbox{or } \phi)$, is taken care of by sampling the decay events simulated with the \texttt{MG5@aMC} code package~\cite{Alwall:2014hca}.
Hence, the total flux of mediator $\Phi_\varphi$ is 
\begin{equation}
    \Phi_\varphi = \sum_i\int_{\rm focus}dE_i d\theta_i \frac{\partial^2 \Phi_i}{\partial E_i \partial \theta_i} {\rm BR}(i\to \varphi)\,, \label{eq:medflux}
\end{equation}
where $i$ runs over all relevant mesons including the neutral and $\Phi_i$ denotes the flux of meson species $i$ for a given protons-on-target (POT). Our $\Phi_i$/POT are normalized to the corresponding numbers reported in \cite{MiniBooNE:2008hfu}. 

In the dark-matter scenario, a produced $V_1$ promptly decays to a dark-matter pair by a 50\% BR as mentioned earlier unless $\epsilon_1$ is too small.\footnote{Note that $g_1$ governing the $V_1\to \chi\chi$ process is not a free parameter but determined by our BR assumption.} If a dark-matter particle passes through the detector fiducial volume, it can go through the process shown in Fig.~\ref{fig:3body}$(b)$. Therefore, the number of signal events $N_S$ in the energy of the $e^+e^-$ pair $E_{ee}$  is
\begin{equation}
   \frac{d N_S}{dE_{ee}} = 2\Phi_{V_1} {\rm BR}(V_1\to 2\chi)A_\chi^{\rm fid} \frac{d \sigma_{\chi N}}{dE_{ee}} N_{T}^{\rm fid} {\rm BR}(V_1\to 2e),
\end{equation}
where $A_\chi^{\rm fid}$, $\sigma_{\chi N}$, and $N_T^{\rm fid}$ are the average probability that $\chi$ travels to the detector fiducial volume, the cross-section of the $\chi N\to \chi' N$ scattering process (see Appendix~\ref{app:dm_scat}), and the number of target nuclei or nucleons in the fiducial volume, respectively.  
Here prefactor 2 accounts for the fact that $V_1$ decays to a dark-matter pair and $\Phi_{V_1}$ is given by Eq.~\eqref{eq:medflux} with $\varphi$ replaced by $V_1$. 

By contrast, in the pseudoscalar scenario,\footnote{For the scalar scenario, the corresponding formalism straightforwardly goes through with $a$ replaced by $\phi$.} a produced $a$ should reach  the detector before it decays and undergo the scattering process shown in Fig.~\ref{fig:3body}$(c)$.
The number of signal events as a function of the photon energy $E_\gamma$ is then expressed as
\begin{equation}
    \frac{d N_S}{dE_\gamma} =\Phi_a A_a^{\rm fid}(1-P_{\rm dec}) \frac{d\sigma_{aN}}{dE_\gamma} N_T^{\rm fid}\,,
\end{equation}
where $A_a^{\rm fid}$, $P_{\rm dec}$, and $\sigma_{aN}$ denote the average probability that $a$ travels to the detector fiducial volume, the average decay probability that $a$ decays before reaching the detector, and the cross-section of the $aN\to \gamma N$ scattering process (see Appendix~\ref{app:dark_prim}).
Again, $\Phi_a$ is given by Eq.~\eqref{eq:medflux} with $\varphi$ replaced by $a$. 

\medskip

\noindent {\bf Fits and discussions}.
We now reproduce the MiniBooNE excess (i.e., residual events) with respect to the two basic experimental observables, visible energy $E_{\rm vis}$ and the angle of visible particle(s) relative to the beam $\cos\theta$ in both neutrino and antineutrino modes, using our models described so far. 
The neutrino and antineutrino mode data are extracted from Ref.~\cite{MiniBooNE:2020pnu} and Ref.~\cite{MiniBooNE:2018esg}, respectively, and a 140~MeV energy threshold, a $<$10$^\circ$ angular separation for $e^+e^-$ pairs, and energy-dependent detection efficiencies~\cite{Patterson:2009ki,Wang:2014nat} are adopted. 
We then use the usual $\chi^2$ function to estimate the goodness of the fit with statistical and systematic uncertainties added by quadrature. Here we approximate systematics in backgrounds, based on the estimates in TABLE I in \cite{MiniBooNE:2020pnu}. 

\begin{figure*}[t]
    \centering
    \includegraphics[width=18cm]{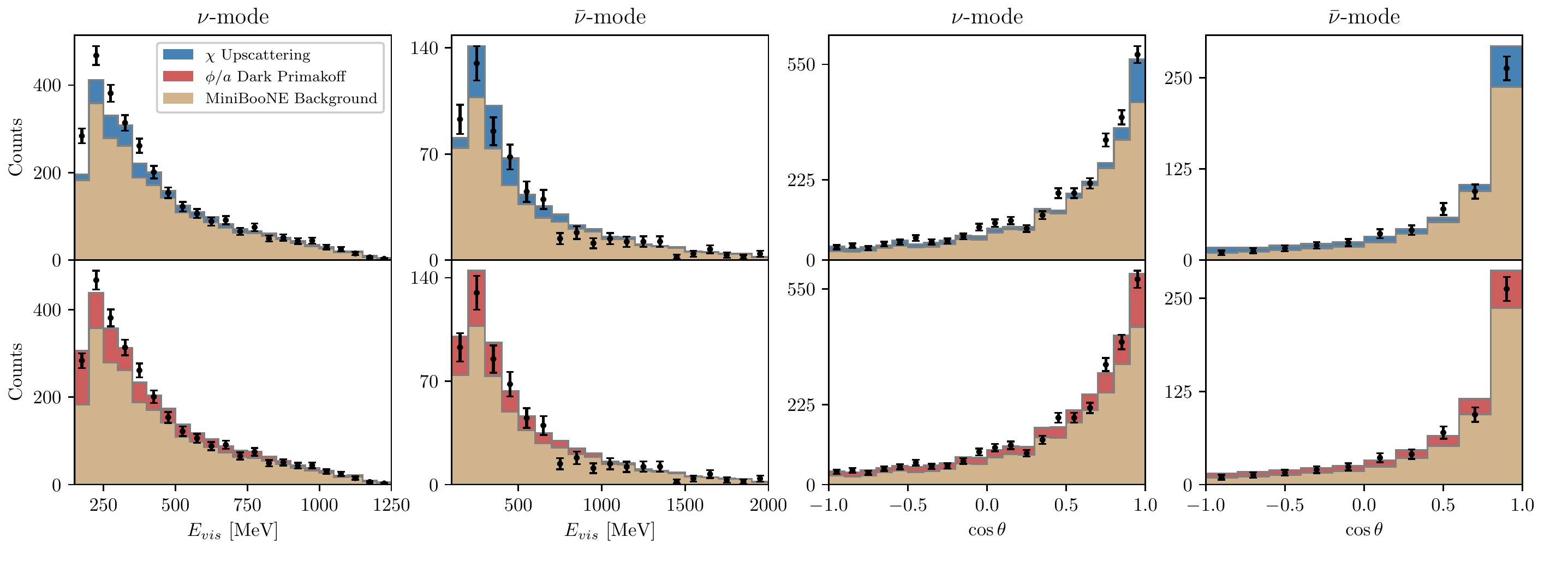}
    \caption{Example fits for the vector-portal dark-matter scenario with double mediators (top panels) and the long-lived scalar scenario  (bottom panels). The parameter values predicted by these fits and the associated $\chi^2$ values are summarized in Table~\ref{tab:fitsummary}. See the text for more detailed discussion.}
    \label{fig:fits}
\end{figure*}

\begin{figure}[h]
    \centering
    \includegraphics[width=0.23\textwidth]{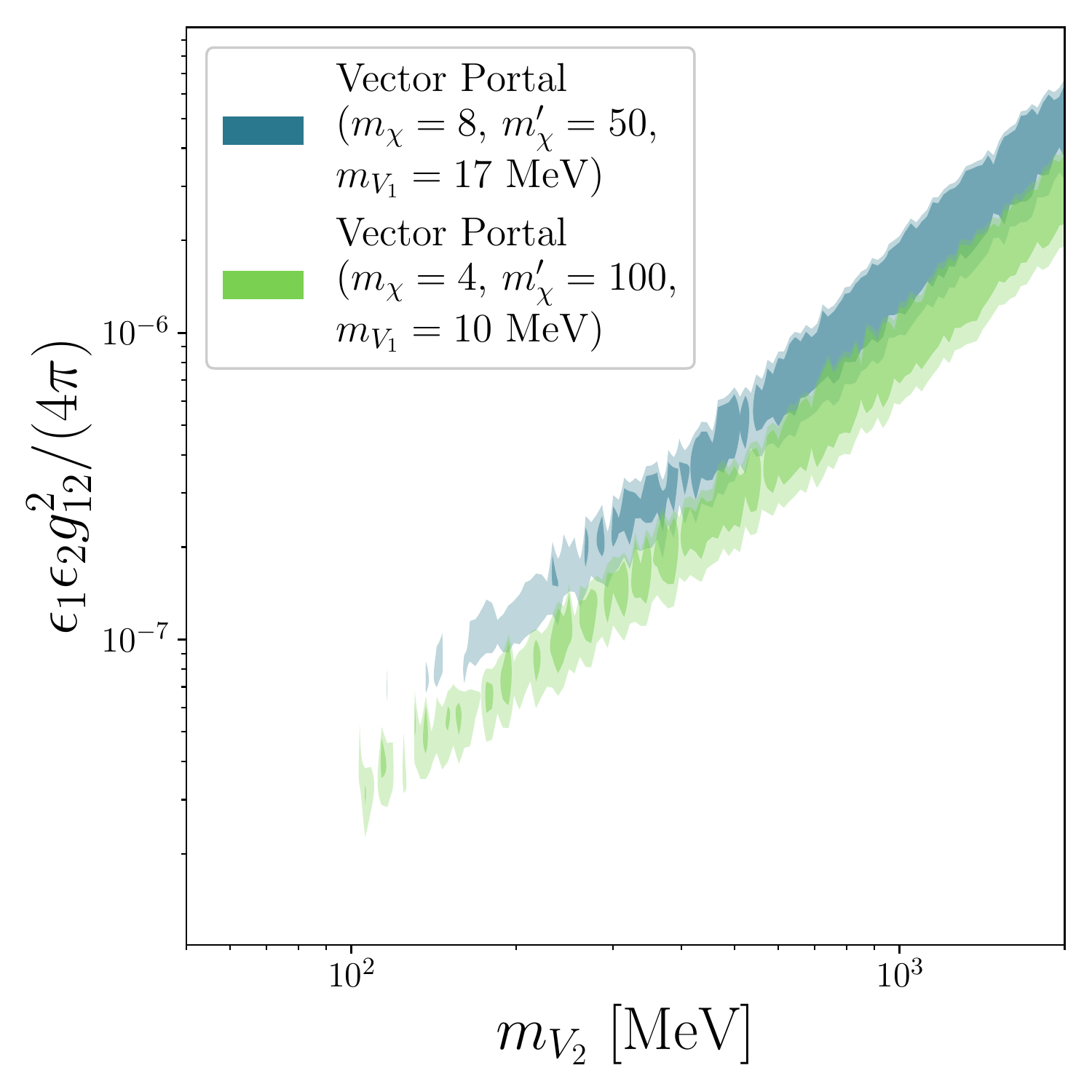}
    \includegraphics[width=0.23\textwidth]{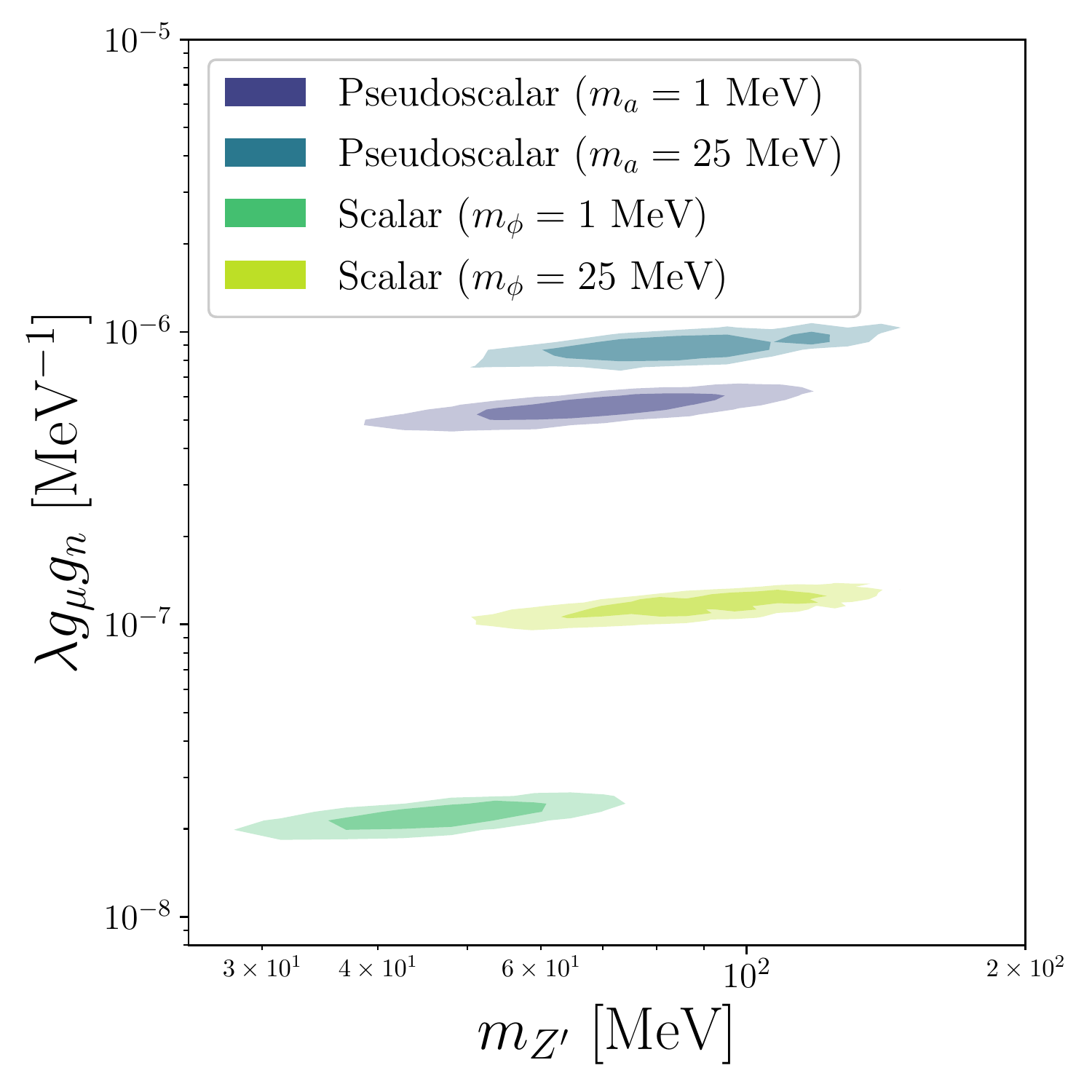}
    \caption{The credible regions for fits to the MiniBooNE excess with model I (left) and model II (right) are shown at 68\% (dark-shaded) and 95\% (light-shaded).
    }
    \label{fig:credible_regions}
\end{figure}

Example fits for model i) and model ii) are displayed in the top panels and the bottom panels of Fig.~\ref{fig:fits}, respectively, and the parameter and $\chi^2$ values associated with these fits are summarized in Table~\ref{tab:fitsummary}. 
The vertical bars describe the statistical uncertainties.
For model i) we assume fermionic $\chi$, $\chi^\prime$ and two different vector mediators $V_1$ and $V_2$, and for model ii) we consider the scalar scenario.
We also find reasonable fits for the long-lived pseudoscalar model as shown in Table~\ref{tab:fitsummary}.
Since MiniBooNE is sensitive to the product of the couplings involved in the scattering process in the detector and the coupling appearing in the charged meson decay, we report the product of those couplings.  
We emphasize that our goal here is to show that there exist a wide range of reasonable parameter sets (see,
for example, Fig.~\ref{fig:credible_regions}), not to spot the best parameter point.

\begin{table}[b]
    \centering
    \resizebox{\columnwidth}{!}{%
    \begin{tabular}{c|c|c|c}
    \hline \hline
    \multicolumn{4}{c}{Vector-portal dark matter} \\
    \hline
    Scenario & $(m_{V_1},m_{V_2},m_\chi,m_{\chi^\prime})$ & $\epsilon_1\epsilon_2 g_2^{\prime 2}/(4\pi)$ & $\chi^2/{\rm dof}$ \\ 
    \hline
    Single & $(17,-,8,40)$~MeV  & $3.6\times 10^{-9}$  & 2.5 (2.9) \\
    Double & $(17,200,8,50)$~MeV  & $1.3\times 10^{-7}$    & 2.2 (2.6) \\
    \hline
    \end{tabular}
    }
    \resizebox{\columnwidth}{!}{%
    \begin{tabular}{c|c|c|c}
    \hline 
    \multicolumn{4}{c}{Long-lived (pseudo)scalar} \\ 
    \hline 
    Scenario & $(m_{Z^\prime},m_{\phi/a})$ & $g_\mu g_n \lambda$ [MeV$^{-1}$] & $\chi^2/{\rm dof}$ \\
    \hline
    Scalar & $(49, 1)$~MeV & $2.2\times 10^{-8}$ & 2.0 (2.1) \\
    Pseudoscalar & $(85, 1)$~MeV & $5.9\times 10^{-7}$ & 2.0 (2.1) \\
    \hline\hline
    \end{tabular}
    }
    \caption{Summary of example fits. 
    In the single-mediator scenario, $m_{V_2}$ is irrelevant, and $\epsilon_2=\epsilon_1$ and $g_2^\prime \to g_1^\prime$. 
    Due to the mass values of the mediators appearing in the scattering process, we fit the data in the limit of nucleon (nucleus) scattering for the double-mediator scenario (the others).
    The $\chi^2$ in the parentheses are the values with statistics only.
    }
    \label{tab:fitsummary}
     \resizebox{\columnwidth}{!}{%
     \begin{tabular}{c|c|c|c|c|c}
    \hline \hline
     Channel& limit  & \multicolumn{2}{c|}{Model i) ($\times 10^{-12}$)} &  \multicolumn{2}{c}{Model ii) ($\times 10^{-8}$)}  \\ \cline{3-6}
     (BR) &($\times10^{-8}$) & Single & Double & $\phi$ & $a$ \\
     \hline
     $K\to\mu\nu_{\mu}V (\phi)$~\cite{NA62:2021bji} & 2000 (300) & 500& 680 & 230 & 100 \\
     $K\to e\nu_{e}\nu\nu$~\cite{ParticleDataGroup:2020ssz} & 6000 & 530& 720 & -- & -- \\
     $K\to\mu(e)\nu_{\mu(e)}ee$~\cite{ParticleDataGroup:2020ssz} & 7.4(2.7)& 500(530) & 680(720) & -- & -- \\
     $\pi\to\mu(e)\nu_{\mu(e)}X$~\cite{PIENU:2021clt} & 600(50)& 0.12(25) & 0.17(34)& 120(--)& 1.1(--)\\
     $\pi\to\mu(e)\nu_{\mu(e)}ee$~\cite{ParticleDataGroup:2020ssz} &  -- (0.37) & 0.12(25) &0.17(34) & -- & -- \\
     \hline \hline 
    \end{tabular}
    }
    \caption{Relevant exotic decays of $\pi^\pm/K^\pm$ and existing upper limits at 90\% CL. $X$ stands for invisibly decaying (massive) bosons. The predicted BRs (third though last columns) are based on the following parameter choices: $(\epsilon_1,\frac{g_1^{\prime2}}{4\pi})\!\simeq\!(6.0\times10^{-5},1)$ for the single-mediator scenario, $(\epsilon_1,\epsilon_2,\frac{g_2^{\prime2}}{4\pi})\!\simeq\!(7.0\times 10^{-5},1.0\times 10^{-4},0.5)$ for the double-mediator scenario, $(g_\mu,g_n,\lambda)\!\simeq\!(5\times10^{-3},10^{-2},4.4\times 10^{-4}\,{\rm MeV}^{-1})$ for the scalar scenario, and $(g_\mu,g_n,\lambda)\!\simeq\!(10^{-2},10^{-2},6.5\cdot10^{-3}\,{\rm MeV}^{-1})$ for the pseudo-scalar scenario. \label{tab:decaywidthlimits}}
\end{table}

We next check if the parameter values for our fits satisfy existing bounds.
First, since our mediators are from the charged-meson decay, the resulting decay widths should agree with exotic $\pi^\pm$/$K^\pm$ decays~\cite{ParticleDataGroup:2020ssz}.
The relevant channels, limits, and BR predictions are summarized in Table~\ref{tab:decaywidthlimits}.\footnote{Be aware that the limits of the $K\to \mu \nu_\mu V(\phi)$ channel are available from 10 MeV, while our (pseudo)scalar has a 1 MeV mass.}

Second, the parameter points should not give rise to any significant number of events in the MiniBooNE off-target mode. Recall that our signal flux mostly originates from charged mesons that are focused, whereas that from neutral mesons are subdominant. In the off-target mode, $\sim\!16$ times smaller POTs were delivered, focusing was unavailable, and the $\pi/K$-decay-in-flight fluxes are smaller by $\sim\!2$ orders of magnitude~\cite{Jordan:2018qiy}.
Our off-target mode simulation suggests a negligible signal flux from charged mesons. 
In the single(double)-mediator scenario of model i), we find that the conventional $\pi^0$ contribution consists of $\sim\!4\,(8)\%$ of signal, potentially yielding $\sim \! 2.4\pm 0.5$ ($4.4\pm0.9$) events in the off-target mode. This agrees with the data within the measurement uncertainty, or could be mitigated by an introduction of a vector mediator without any significant coupling to the first generation of quarks.
For the (pseudo)scalar scenario, we find that $\sim\!1$ event is expected in the off-target mode. 

Third, we check limits from other dark-sector particle search experiments. Since we assume that the dark photon(s) in model i) are half-(in)visible, the limits of both invisibly- and visibly-decaying dark photons are relevant modulo the BR. 
For the parameter points in Table~\ref{tab:fitsummary}, the leading limits are $\epsilon\approx0.074~(1.2)\times 10^{-3}$ for a 17 (200) MeV dark photon from NA64 (invisible dark photon decay)~\cite{Banerjee:2019pds} and $\epsilon\approx0.092~(1.4)\times 10^{-3}$ for a 17 (200) MeV dark photon from E141~\cite{Riordan:1987aw} and BaBar (visible dark photon decay)~\cite{BaBar:2014zli}, which are reinterpreted again under the assumption that the BRs of the visibly- and invisibly-decaying modes are 50\% and 50\%. 
Our parameter choices mentioned in Table~\ref{tab:decaywidthlimits} are not excluded by these experiments.
Regarding the recent COHERENT result~\cite{coherentrecent}, we find that it is insensitive to our vector-portal dark-matter model. COHERENT would observe signal events through the coherent elastic scattering of our dark matter. 
However, as mentioned earlier, coupling $g_1$ is roughly proportional to $\epsilon_1$ for our dark photon to be half-(in)visible, and thus the resulting scattering cross-section at detection would be too small.\footnote{By contrast, in COHERENT, $\alpha_D$ associated with $g_1$ is assumed to be 0.5.} 
Regarding Model ii), we emphasize that the limits from other search experiments are highly model-dependent, while we have not resorted to a particular model. %In Appendix~D, 
For interested readers, we provide  model examples  accommodating \eqref{eq:lagrangian_sp} in Appendix~\ref{app:scalarmodel} and explain how the related limits are evaded.
A detailed study of UV models will be presented in an upcoming publication.

Fourth, we find that our models are consistent with the recent MicroBooNE results \cite{MicroBooNE:2021zai}. Compared to MiniBooNE,  it is based on $\sim\!3$ times smaller POT~\cite{MiniBooNE:2020pnu,MicroBooNE:2021zai}, $\sim\!6$ times smaller fiducial volume~\cite{MiniBooNE:2020pnu,MicroBooNE:2021gfj}, and $\sim\!3$ times smaller photon detection efficiency~\cite{Wang:2014nat,MicroBooNE:2021zai}.  For a coherent scattering process the liquid argon of MicroBooNE gets a $\sim\!3$ factor increase relative to the mineral oil of MiniBooNE. This gives a net reduction factor of $\sim\!18$.  Given that MiniBooNE observed 320 excess events below 300 MeV visible energy \cite{MiniBooNE:2020pnu}, then the models presented here with coherent scattering predict MicroBooNE  would expect $\sim\!18$ event excess in the $1\gamma0p$ analysis at low energy.  It is interesting that they report a 18 event (2.7$\sigma$) excess for the $1\gamma0p$ sample in the 200-250 MeV visible energy bin \cite{MicroBooNE:2021zai}, demonstrating consistency with our predictions.  Furthermore, the more recent MicroBooNE result demonstrates that less than approximately 50\% of the MiniBooNE excess can come from $\nu_e$ charged-current scattering \cite{MicroBooNE:2021rmx}, suggesting that the majority of the excess is not from intrinsic $\nu_e$ backgrounds or sterile neutrino oscillations.  This further bolsters the building evidence that the MiniBooNE excess is real and unexplained and it is potentially from a new source, such as the models presented here. 

 We finally remark that in the scalar and pseudoscalar scenarios, couplings such as $a F \tilde{F}$ are negligible in the fit to ensure decays $a \to \gamma \gamma$ do not contribute too much to the forward events that would create too much of an excess in the forward-most cosine bin.

\medskip 

\noindent {\bf Prospects at Neutrino Experiments}. 
Experiments that utilize proton beams and the induced neutrino fluxes should also be sensitive to new physics emerging in the charged mesons that we have used to explain the MiniBooNE excess.\footnote{the long-lived (pseudo)scalar model can potentially explain the LSND excess~\cite{LSND:2001aii} by initiating a neutron ejection from the inelastic interaction of $Z^\prime$ with a nucleus at the detector and subsequent neutron capture would produce the 2.2 MeV photon signal. However, since the proton beam energy of LSND is just 800 MeV, dark matter $\chi$ in model i) would not upscatter to $\chi^\prime$ that can create an electron-like signature through its decay, and thus the excess would not be explained.} 
Obviously, the experiments (e.g., SBND) using the same beam as for MiniBooNE can allow for checking the idea and predictions presented here~\cite{us}.
Those using a higher-energy beam (e.g., DUNE/FASER/ICARUS-NuMI)
can also test the proposed models, based on similar signal production mechanisms and detection channels. 
In particular, since charged mesons are produced with a larger boost, their decay products are more likely to lie in the forward region and thus a more signal flux can enter the detector. 
\footnote{By contrast, this  excess is not observable at CHARM-II, MINER$\nu$A, and T2K after considering their detector sizes, $\pi^\pm$-production rates, and target-to-detector distances altogether.}

Likewise, the low-energy high-intensity beam-based experiments (e.g., CCM/COHERENT/JSNS$^2$) may have interesting opportunities.
In these experiments, no beam focusing is available, so contributions from the charged mesons would be weakened, whereas those from the neutral mesons (e.g., $\pi^0$) would relatively stand out. 
Nevertheless, higher beam intensity and close proximity of the detector to the beam target would allow for a rather significant number of signal events. 
Especially, only the scattering-based MiniBooNE solutions involving $\mathcal{O}(1-10)$~MeV mediators can be directly checked in these experiments.
In the case of model i), signal detection via the dark-matter upscattering would not be available, as the associated dark-matter particles are not energetic enough to create a heavier state, $\chi^\prime$. By contrast, given lower energy thresholds, coherent dark-matter elastic scattering can be promising for testing our dark-matter model in the double-mediator scenario, as $g_2$ can be sizable enough (unlike $g_1$) for ongoing/upcoming beam-based CE$\nu$NS experiments to observe signal events. 
On the other hand, in the case of model ii), $Z^\prime$ can arise from the $\pi^0$ decay and decay into a (pseudo)scalar through $aF^\prime \tilde{F}$ or $\phi F^\prime F$.  
For signal detection, the same strategy is applicable.
Again thanks to the lower energy thresholds, both the outgoing photon and the nuclear recoil can be recorded simultaneously (i.e., fully visible), allowing us to infer properties (e.g., mass) of the incoming particle more accurately.  
Finally, we remark that JSNS$^2$ can provide a unique opportunity especially if the MiniBooNE signal is only or dominantly sourced by charged kaons.

\medskip 

\section*{Acknowledgement}
We thank Yue Zhao for useful and insightful discussions. 
The work of BD, DK, and AT are supported by the DOE Grant No. DE-SC0010813. We acknowledge the support of the Department of Energy Office of Science, Los Alamos National Laboratory LDRD funding, and funding from the National Laboratories Office at Texas A\&M. We acknowledge that portions of this research were conducted with the advanced computing resources provided by Texas A\&M High Performance Research Computing.

\medskip

\onecolumngrid
\appendix
\section{Charged Meson Three-Body Decay} \label{app:decay_width}

\begin{figure}[t]
    \centering
    \includegraphics[width=0.49\textwidth]{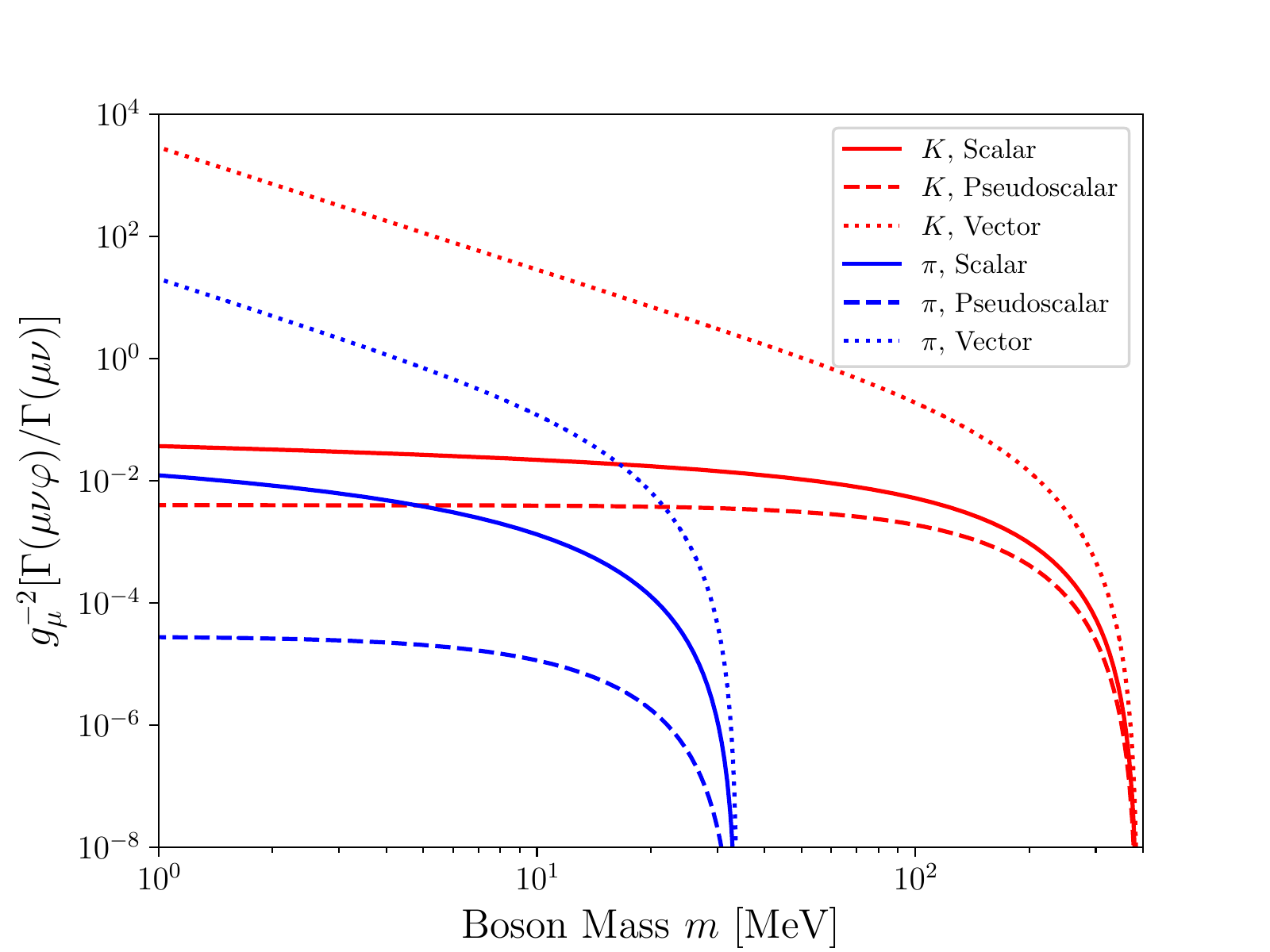}
    \includegraphics[width=0.49\textwidth]{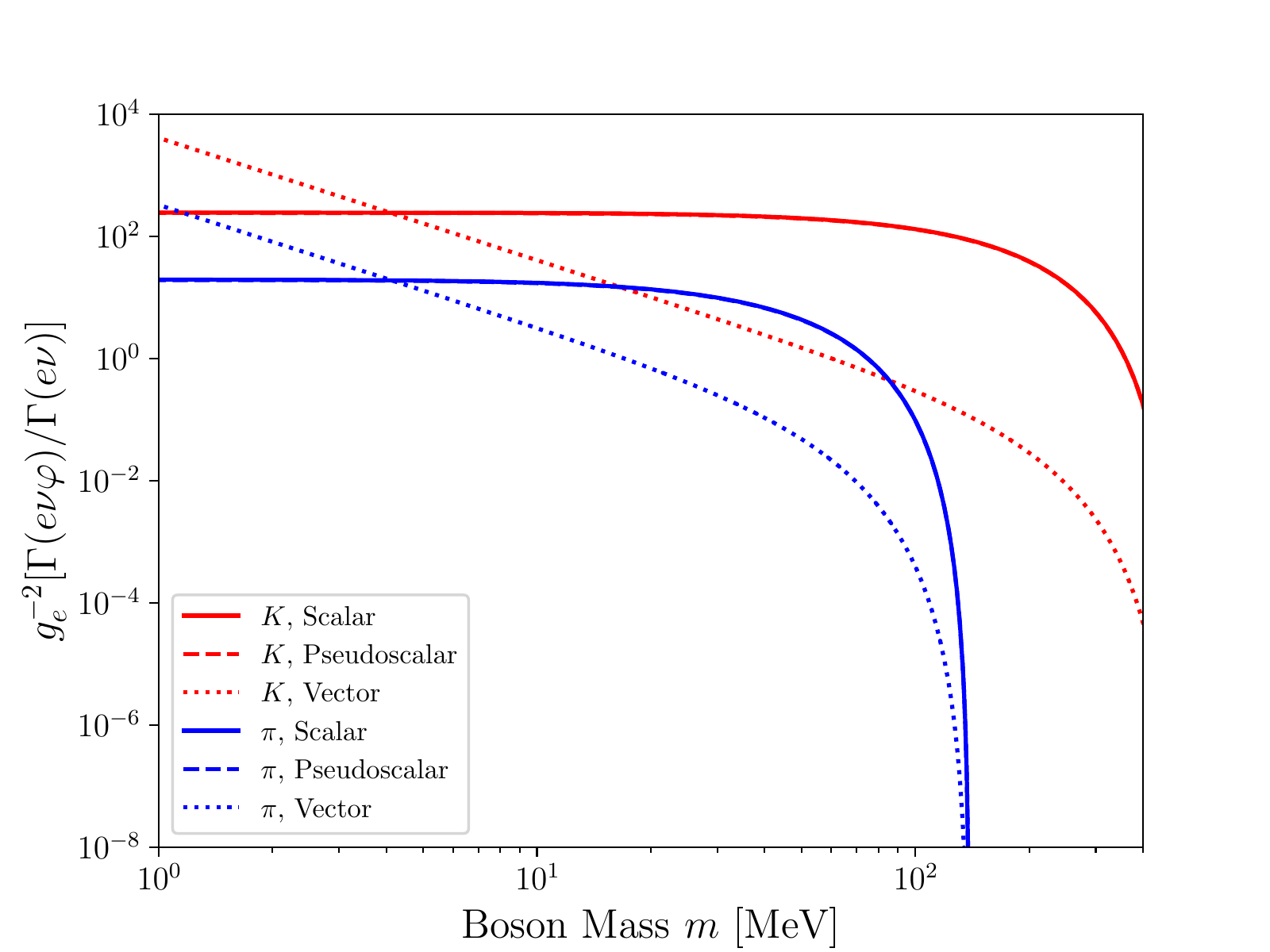}
    \caption{Branching ratios for the three-body production of scalars, pseudoscalars, and vectors ($\varphi = \phi, a, V$) whose mass is $m$ via the decay of a charged meson with mass $M (= m_{\pi}, m_{K})$, relative to their corresponding two-body decay. The charged meson decays involve a muon (left panel) or an electron (right panel). }
    \label{fig:br}
\end{figure}

Production of a boson $\varphi$ (vector/scalar/pseudoscalar) from the decay of charged mesons has been studied in literature, e.g., Refs.~\cite{Carlson:2012pc, Altmannshofer:2019yji, Krnjaic:2019rsv}; the amplitudes we consider for new scalar, pseudoscalar, and vector bosons are depicted diagrammatically in Fig.~1 in the main text. The noteworthy quality of these decay modes is the large phase space available to them, relative to ordinary charged meson two-body decays. 
Unlike the two-body decay, which experiences a phase space suppression due to the selection of only one possible helicity combination for the outgoing fermions, the three-body final state's extra degree of freedom can alleviate this suppression. This is manifested in the amplitude through the spin sum of $\sim \nu^\dagger \mu_\uparrow$, which ordinarily selects out only $\nu_\downarrow \mu_\uparrow$ since the muon and muon neutrino are forced to be back-to-back in the meson rest frame. In the three-body final state, however, the momentum of the new boson gives the outgoing fermions a free separation angle that does not simply collapse the spinor contraction. 
We see this in Fig.~\ref{fig:br}.
For the muonic decay case (left panel), the phase-space suppression for scalars is only about a factor of $10^{-2}$, and in the case of vectors, we see a phase-space enhancement. For pseudoscalars, however, we see a stronger suppression because of their odd-parity nature giving rise to a destructive interference between helicity final states, manifested by a term like $-\nu^\dagger m_\mu m_\pi \mu_\uparrow$ in the case of pion decay or $-\nu^\dagger m_\mu m_K \mu_\uparrow$ in the case of kaon decay. 
However, this effect becomes smaller in the electronic decay case (right panel). Moreover, the larger mass gap between the meson and the electron allows for more enhancement relative to the $\pi/K\to e\nu_e$ decay.

\section{Upscattering of Dark Matter \label{app:dm_scat} } 

The matrix element sqaured for the dark-matter coherent upscattering process $\chi N \to \chi^\prime N$ has the form of
\begin{equation}
    \left.\overline{|\mathcal{M}|^2}\right|_{\rm nucleus}=\frac{(e\epsilon_2 g^\prime_2)^2}{\left( t-m_{V_2}^2\right)^2}\mathcal{M}_0\,,
\end{equation}
where we assume that dark photon $V_2$ is exchanged in the $t$-channel and factorize the model-dependent terms to $\mathcal{M}_0$. Here $m_N$ is the mass of the target nucleus, and $t\equiv (p_\chi-p_{\chi^\prime})^2=2m_N(E_{\chi^\prime}-E_\chi)$ is a usual Mandelstam variable.
For the fermionic nucleus (and nucleon as well) target with scalar and fermionic dark matter, $\mathcal{M}_0$ are respectively given as follows~\cite{Giudice:2017zke}:
\begin{eqnarray}
    \mathcal{M}_{0, FS} &=& 8m_N\left(2m_N E_\chi E_{\chi^\prime}+m_\chi^2 E_{\chi^\prime}-m_{\chi^\prime}^2 E_\chi\right), \\
    \mathcal{M}_{0, FF} &=& 8m_N \left[m_N^2(E_{\chi^\prime}-E_\chi)+m_N\left(E_\chi^2 +E_{\chi^\prime}^2\right)-\frac{(m_{\chi^\prime}-m_\chi)^2}{2}\left(E_{\chi^\prime}-E_\chi+m_N \right)+m_\chi^2 E_{\chi^\prime}-m_{\chi^\prime}^2 E_\chi \right]\,.
\end{eqnarray}
The expressions corresponding to the scalar nucleus target are 
\begin{eqnarray}
    \mathcal{M}_{0, SS} &=& 4m_N^2 \left(E_\chi+E_{\chi^\prime} +\frac{m_\chi^2-m_{\chi^\prime}^2}{2m_N} \right)^2,\\
    \mathcal{M}_{0, SF} &=& 8m_N^3(E_{\chi^\prime}-E_\chi)+4m_N^2\left\{4E_\chi E_{\chi^\prime}-(m_{\chi^\prime}-m_\chi)^2 \right\} \nonumber \\
    &-&2m_N(m_{\chi^\prime}-m_\chi)\left\{E_{\chi^\prime}(m_{\chi^\prime}+3m_\chi)+E_\chi(3m_{\chi^\prime}+m_\chi) \right\}+\left(m_{\chi^\prime}^2-m_\chi^2 \right)^2.
\end{eqnarray}

The differential cross-section in the laboratory frame for the (generic) scattering process between dark matter and a nucleus~\cite{Dutta:2019nbn,Dutta:2020vop} is
\begin{equation}
    \left.\frac{d\sigma_{\chi N}}{dE_N}\right|_{\rm nucleus} = \frac{Q_{\rm eff}^2}{32\pi m_N p_\chi^2} F_N^2(t)\left.\overline{|\mathcal{M}|^2}\right|_{\rm nucleus}\,,
\end{equation}
where the nuclear form factor is encoded in $F_N(t)$ for which we adopt the Helm parametrization~\cite{Helm:1956zz} in our study. $Q_{\rm eff}$ denotes an effective gauge charge for a given nucleus: for example, atomic number $Z$ for the usual dark photon models and atomic mass number $A$ for the $B-L$ models. 
Finally, $E_N$ is the total energy of the recoiling nucleus and $\overline{|\mathcal{M}|^2}$ can be written in terms of $E_N=E_\chi+m_N-E_{\chi^\prime}$. 
The kinematically allowed upper and lower limits of $E_N$, denoted by $E_N^+$ and $E_N^-$, are
\begin{equation}
    E_N^\pm = \frac{s+m_N^2-m_{\chi^\prime}^2}{2\sqrt{s}}\frac{E_\chi+m_N}{\sqrt{s}}\pm \frac{\lambda^{1/2}\left(s,m_N^2,m_{\chi^\prime}^2\right)}{2\sqrt{s}}\frac{p_\chi}{\sqrt{s}}\,, \label{eq:kinrange}
\end{equation}
where the center-of-mass energy squared $s$ and the kinematic triangular function $\lambda$ are defined as $s=m_N^2+m_\chi^2+2m_N E_\chi$ and $\lambda(x,y,z)=(x-y-z)^2-4yz$.
In the non-relativistic limit, which is the case of nuclear recoils, one may replace $E_N$ by $m_N+E_r$ with $E_r$ being the recoil kinetic energy. 

When it comes to the nucleon scattering, the differential cross-section has a similar form with nucleon form factors incorporated into the modified matrix element squared:
\begin{equation}
    \left.\frac{d\sigma_{\chi N}}{dE_N}\right|_{\rm nucleon}= \frac{Q_N^2}{32\pi m_N p_\chi^2} \left.\overline{|\mathcal{M}|^2}\right|_{\rm nucleon}\,,
\end{equation}
where $m_N$ and $Q_N$ are the mass and the gauge charge of the target nucleon. These two quantities are model-dependent; for example, in models of dark photon, the proton scattering is allowed so that $m_N=m_p$ and $Q_N=+1$. 
The kinematically allowed range of $E_N$ is again defined by Eq.~\eqref{eq:kinrange}.
In the matrix element calculation, the vertex associated with the nucleon current, $\bar{u}_N \gamma^\mu u_N$, can be replaced by~\cite{Borie:2012tu} 
\begin{equation}
    \bar{u}_N\left(F_1(t)\gamma^\mu +\kappa F_2(t)\frac{i\sigma^{\mu\nu}q_\nu}{2m_N} \right)u_N\,, \label{eq:gencurrent}
\end{equation}
where $t=q^2$ and $\sigma^{\mu\nu}=\frac{i}{2}[\gamma^\mu,\gamma^\nu]$ and where the anomalous magnetic moment $\kappa=1.79$ and $-1.91$ for the proton and neutron.
Now plugging \eqref{eq:gencurrent} into the scattering amplitude, one can find that the matrix element squared for the fermionic dark matter~\cite{Kim:2016zjx} is
\begin{equation}
    \left.\overline{|\mathcal{M}|^2}\right|_{\rm nucleon}=\frac{(e\epsilon_2 g^\prime_2)^2}{\left( t-m_{V_2}^2\right)^2}\left[\mathcal{M}_{0,FF}\left(F_1+\kappa F_2\right)^2+\mathcal{M}_{1,FF}\left\{-\left(F_1+\kappa F_2\right) \kappa F_2+\frac{E_\chi-E_{\chi^\prime}+2m_N}{4m_N}(\kappa F_2)^2 \right\}\right],
\end{equation}
where $\mathcal{M}_{1,FF}$ has the form of~\cite{Kim:2016zjx}
\begin{equation}
    \mathcal{M}_{1,FF}= 8m_N^2\left[\left(E_\chi+E_{\chi^\prime}-\frac{m_{\chi^\prime}^2-m_\chi^2}{2m_N} \right)^2+ \left(E_\chi-E_{\chi^\prime}+2m_N \right) \left(E_{\chi^\prime}-E_{\chi}-\frac{(m_{\chi^\prime}-m_\chi)^2}{2m_N} \right) \right].
\end{equation}

For the proton target, the form factors have relationship with the Sachs electric/magnetic form factors $G_E$/$G_M$ such that 
\begin{equation}
    F_1 = \frac{G_E-\frac{t}{4m_p^2}}{1-\frac{t}{4m_p^2}},~~~\kappa F_2=\frac{G_M-G_E}{1-\frac{t}{4m_p^2}}.
\end{equation}
The Rosenbluth formula in combination with experimental measurements
suggests the following dipole approximation
\begin{equation}
    G_E = \frac{G_M}{\mu_p}=\left(1-\frac{t}{0.71 {\rm GeV}^2}\right)^{-2}
\end{equation}
up to about $t\sim -10 {\rm GeV}^2$ with $\mu_p=\frac{(1+\kappa_p)e}{2m_p}=2.79$~\cite{Qattan:2004ht}. The form factors also have the following relations:
\begin{equation}
    G_M=F_1+\kappa_p F_2,~~~ G_E=F_1+\frac{t}{4m_p^2}\kappa_p F_2.
\end{equation}

\section{``Dark Primakoff'' Scattering}\label{app:dark_prim}
We now consider the scattering of a pseudoscalar through $a(k) N(p) \to \gamma(k^\prime) N(p^\prime)$ or scalar through $\phi(k) N(p) \to \gamma(k^\prime) N(p^\prime)$ via exchange of a massive vector $Z^\prime$ that interacts with a nucleon ($N = p,n$) in a target (fermionic) nucleus of mass $m_N$. The matrix element for a pseudoscalar $a$ is
\begin{equation}
     i\mathcal{M} = \Bar{u}(p^\prime) (-i g_n \gamma^\mu) u(p) \bigg(\dfrac{-i(g_{\mu\nu} - \frac{q_\mu q_\nu}{m_{Z^\prime}^2})}{q^2 - m_{Z^\prime}^2} \bigg) (i \lambda \epsilon^{\nu\rho\alpha\beta} q_\alpha k^\prime_\beta) \varepsilon^*_\rho (k^\prime)
\end{equation}
which yields
\begin{equation}
    \overline{\mid \mathcal{M}\mid^2} = \frac{g_n^2 \lambda^2 t \left\{2 m_N^2 \left(m_a^2-2 s-t\right)+2 m_N^4-2 m_a^2 (s+t)+m_a^4+2 s^2+2 s t+t^2\right\}}{2 \left(t-m_{Z^\prime}^2\right)^2} ,
    \label{eq:m2ps}
\end{equation}
The differential scattering cross-section is then expressed with a nuclear form factor $F_N(t)$ in the following form,
\begin{equation}
    \dfrac{d\sigma}{dt} = \dfrac{Z^2}{16\pi\{s-(m_a + m_N)^2\}\{s-(m_a - m_N)^2\}} \overline{\mid \mathcal{M}\mid^2} F_N^2(t),
\end{equation}
where again $Z$ is the atomic number of the target nucleus. 
The matrix element for a scalar $\phi$ is
\begin{equation}
    \mathcal{M} = \Bar{u}(p^\prime) (-i g_n \gamma^\mu) u(p) \bigg(\dfrac{-i(g_{\mu\nu} - \frac{q_\mu q_\nu}{m_{Z^\prime}^2})}{q^2 - m_{Z^\prime}^2} \bigg) \bigg(i \lambda (q^\rho k^{\prime \nu} - (k^\prime \cdot q) g^{\rho\nu})\bigg) \varepsilon^*_\rho (k^\prime)
\end{equation}
which yields the same squared matrix element as in Eq.~\eqref{eq:m2ps} with the replacement $m_a \to m_\phi$.

\section{Examples of  Models for Scattering and Limits }\label{app:scalarmodel} 
For a given model,  both quark and lepton couplings of the mediators (scalars/pseudoscalars/vector) depending on the model requirement need to be considered for the pion three-body decay along with various scattering  possibilities via different mediators. 
Furthermore, the (pseudo)scalar couplings to quarks and leptons can be different (since these are Yukawa couplings and are free to choose in any model). With all these possibilities, there can be many simple models [with vector and (pseudo)scalar mediators] which can fit the excess while explaining all other constraints. Both long-lived vector and (pseudo)scalar mediator scenarios produce similar fits of the MiniBooNE anomaly. Suppose we use a long-lived vector mediator scenario for the MB fit where the vector mediator emits from the charged pion decay and it gets scattered into photon after exchanging (pseudo)scalar particle with the nucleus. This scenario requires a smaller coupling combination, i.e., $g^V_{\pi} g_n\lambda\sim 10^{-(9-10)}$~MeV$^{-1}$ for $m_V\sim\mathcal{O}(10)$ MeV and (pseudo)scalar mass $\sim \mathcal{O}(100)$ MeV compared to the scenario where a (pseudo)scalar particle emerges from the charged pion and then produces a photon by exchanging a vector boson with  the nucleus. In this long-lived vector mediator scenario, one can easily choose the vector mediator coupling to quark $\lesssim 10^{-3}$~\cite{Holst:2021lzm} which would satisfy muon $g$-2 constraint (even if the same gauge coupling appears for muons).  One can further simplify this scenario by replacing the (pseudo)scalar exchanged with the nucleus  by $\pi^0$. 
In this simplified setup one does not even need any new scalar with  $\lambda\sim 10^{-3} g^V$~MeV$^{-1}$ 
and $g_n=g_{\pi NN}\sim 1$~\cite{deSwart:1997ep}. 
Considering the above estimations we find that, the  if we choose a particular model which explains e.g., both  $g$-2 and  $R_{K^{(\ast)}}$  anomalies in the context of an anomaly-free $U(1)_{T3R}$ extension model presented in Ref.~\cite{Dutta:2021afo} after obeying not only the cosmological and astrophysical constraints but also the $B_s$-$\bar{B}_{s}$ mixing constraint, the model can explain the excess. Similarly one can choose other types of $U(1)$ model to explain the excess.

\bibliography{main}

%merlin.mbs apsrev4-1.bst 2010-07-25 4.21a (PWD, AO, DPC) hacked
%Control: key (0)
%Control: author (72) initials jnrlst
%Control: editor formatted (1) identically to author
%Control: production of article title (-1) disabled
%Control: page (0) single
%Control: year (1) truncated
%Control: production of eprint (0) enabled
\begin{thebibliography}{93}%
\makeatletter
\providecommand \@ifxundefined [1]{%
 \@ifx{#1\undefined}
}%
\providecommand \@ifnum [1]{%
 \ifnum #1\expandafter \@firstoftwo
 \else \expandafter \@secondoftwo
 \fi
}%
\providecommand \@ifx [1]{%
 \ifx #1\expandafter \@firstoftwo
 \else \expandafter \@secondoftwo
 \fi
}%
\providecommand \natexlab [1]{#1}%
\providecommand \enquote  [1]{``#1''}%
\providecommand \bibnamefont  [1]{#1}%
\providecommand \bibfnamefont [1]{#1}%
\providecommand \citenamefont [1]{#1}%
\providecommand \href@noop [0]{\@secondoftwo}%
\providecommand \href [0]{\begingroup \@sanitize@url \@href}%
\providecommand \@href[1]{\@@startlink{#1}\@@href}%
\providecommand \@@href[1]{\endgroup#1\@@endlink}%
\providecommand \@sanitize@url [0]{\catcode `\\12\catcode `\$12\catcode
  `\&12\catcode `\#12\catcode `\^12\catcode `\_12\catcode `\%12\relax}%
\providecommand \@@startlink[1]{}%
\providecommand \@@endlink[0]{}%
\providecommand \url  [0]{\begingroup\@sanitize@url \@url }%
\providecommand \@url [1]{\endgroup\@href {#1}{\urlprefix }}%
\providecommand \urlprefix  [0]{URL }%
\providecommand \Eprint [0]{\href }%
\providecommand \doibase [0]{http://dx.doi.org/}%
\providecommand \selectlanguage [0]{\@gobble}%
\providecommand \bibinfo  [0]{\@secondoftwo}%
\providecommand \bibfield  [0]{\@secondoftwo}%
\providecommand \translation [1]{[#1]}%
\providecommand \BibitemOpen [0]{}%
\providecommand \bibitemStop [0]{}%
\providecommand \bibitemNoStop [0]{.\EOS\space}%
\providecommand \EOS [0]{\spacefactor3000\relax}%
\providecommand \BibitemShut  [1]{\csname bibitem#1\endcsname}%
\let\auto@bib@innerbib\@empty
%</preamble>
\bibitem [{\citenamefont {Aguilar-Arevalo}\ \emph
  {et~al.}(2009{\natexlab{a}})\citenamefont {Aguilar-Arevalo} \emph
  {et~al.}}]{MiniBooNE:2008yuf}%
  \BibitemOpen
  \bibfield  {author} {\bibinfo {author} {\bibfnamefont {A.~A.}\ \bibnamefont
  {Aguilar-Arevalo}} \emph {et~al.} (\bibinfo {collaboration} {MiniBooNE}),\
  }\href {\doibase 10.1103/PhysRevLett.102.101802} {\bibfield  {journal}
  {\bibinfo  {journal} {Phys. Rev. Lett.}\ }\textbf {\bibinfo {volume} {102}},\
  \bibinfo {pages} {101802} (\bibinfo {year} {2009}{\natexlab{a}})},\ \Eprint
  {http://arxiv.org/abs/0812.2243} {arXiv:0812.2243 [hep-ex]} \BibitemShut
  {NoStop}%
\bibitem [{\citenamefont {Aguilar-Arevalo}\ \emph
  {et~al.}(2018{\natexlab{a}})\citenamefont {Aguilar-Arevalo} \emph
  {et~al.}}]{MiniBooNE:2018esg}%
  \BibitemOpen
  \bibfield  {author} {\bibinfo {author} {\bibfnamefont {A.~A.}\ \bibnamefont
  {Aguilar-Arevalo}} \emph {et~al.} (\bibinfo {collaboration} {MiniBooNE}),\
  }\href {\doibase 10.1103/PhysRevLett.121.221801} {\bibfield  {journal}
  {\bibinfo  {journal} {Phys. Rev. Lett.}\ }\textbf {\bibinfo {volume} {121}},\
  \bibinfo {pages} {221801} (\bibinfo {year} {2018}{\natexlab{a}})},\ \Eprint
  {http://arxiv.org/abs/1805.12028} {arXiv:1805.12028 [hep-ex]} \BibitemShut
  {NoStop}%
\bibitem [{\citenamefont {Aguilar-Arevalo}\ \emph
  {et~al.}(2021{\natexlab{a}})\citenamefont {Aguilar-Arevalo} \emph
  {et~al.}}]{MiniBooNE:2020pnu}%
  \BibitemOpen
  \bibfield  {author} {\bibinfo {author} {\bibfnamefont {A.~A.}\ \bibnamefont
  {Aguilar-Arevalo}} \emph {et~al.} (\bibinfo {collaboration} {MiniBooNE}),\
  }\href {\doibase 10.1103/PhysRevD.103.052002} {\bibfield  {journal} {\bibinfo
   {journal} {Phys. Rev. D}\ }\textbf {\bibinfo {volume} {103}},\ \bibinfo
  {pages} {052002} (\bibinfo {year} {2021}{\natexlab{a}})},\ \Eprint
  {http://arxiv.org/abs/2006.16883} {arXiv:2006.16883 [hep-ex]} \BibitemShut
  {NoStop}%
\bibitem [{\citenamefont {Brdar}\ and\ \citenamefont
  {Kopp}(2021)}]{Brdar:2021cgb}%
  \BibitemOpen
  \bibfield  {author} {\bibinfo {author} {\bibfnamefont {V.}~\bibnamefont
  {Brdar}}\ and\ \bibinfo {author} {\bibfnamefont {J.}~\bibnamefont {Kopp}},\
  }\href@noop {} {\  (\bibinfo {year} {2021})},\ \Eprint
  {http://arxiv.org/abs/2109.08157} {arXiv:2109.08157 [hep-ph]} \BibitemShut
  {NoStop}%
\bibitem [{\citenamefont {Abratenko}\ \emph
  {et~al.}(2021{\natexlab{a}})\citenamefont {Abratenko} \emph
  {et~al.}}]{MicroBooNE:2021zai}%
  \BibitemOpen
  \bibfield  {author} {\bibinfo {author} {\bibfnamefont {P.}~\bibnamefont
  {Abratenko}} \emph {et~al.} (\bibinfo {collaboration} {MicroBooNE}),\
  }\href@noop {} {\  (\bibinfo {year} {2021}{\natexlab{a}})},\ \Eprint
  {http://arxiv.org/abs/2110.00409} {arXiv:2110.00409 [hep-ex]} \BibitemShut
  {NoStop}%
\bibitem [{\citenamefont {Sorel}\ \emph {et~al.}(2004)\citenamefont {Sorel},
  \citenamefont {Conrad},\ and\ \citenamefont {Shaevitz}}]{Sorel:2003hf}%
  \BibitemOpen
  \bibfield  {author} {\bibinfo {author} {\bibfnamefont {M.}~\bibnamefont
  {Sorel}}, \bibinfo {author} {\bibfnamefont {J.~M.}\ \bibnamefont {Conrad}}, \
  and\ \bibinfo {author} {\bibfnamefont {M.}~\bibnamefont {Shaevitz}},\ }\href
  {\doibase 10.1103/PhysRevD.70.073004} {\bibfield  {journal} {\bibinfo
  {journal} {Phys. Rev. D}\ }\textbf {\bibinfo {volume} {70}},\ \bibinfo
  {pages} {073004} (\bibinfo {year} {2004})},\ \Eprint
  {http://arxiv.org/abs/hep-ph/0305255} {arXiv:hep-ph/0305255} \BibitemShut
  {NoStop}%
\bibitem [{\citenamefont {Karagiorgi}\ \emph {et~al.}(2009)\citenamefont
  {Karagiorgi}, \citenamefont {Djurcic}, \citenamefont {Conrad}, \citenamefont
  {Shaevitz},\ and\ \citenamefont {Sorel}}]{Karagiorgi:2009nb}%
  \BibitemOpen
  \bibfield  {author} {\bibinfo {author} {\bibfnamefont {G.}~\bibnamefont
  {Karagiorgi}}, \bibinfo {author} {\bibfnamefont {Z.}~\bibnamefont {Djurcic}},
  \bibinfo {author} {\bibfnamefont {J.~M.}\ \bibnamefont {Conrad}}, \bibinfo
  {author} {\bibfnamefont {M.~H.}\ \bibnamefont {Shaevitz}}, \ and\ \bibinfo
  {author} {\bibfnamefont {M.}~\bibnamefont {Sorel}},\ }\href {\doibase
  10.1103/PhysRevD.81.039902} {\bibfield  {journal} {\bibinfo  {journal} {Phys.
  Rev. D}\ }\textbf {\bibinfo {volume} {80}},\ \bibinfo {pages} {073001}
  (\bibinfo {year} {2009})},\ \bibinfo {note} {[Erratum: Phys.Rev.D 81, 039902
  (2010)]},\ \Eprint {http://arxiv.org/abs/0906.1997} {arXiv:0906.1997
  [hep-ph]} \BibitemShut {NoStop}%
\bibitem [{\citenamefont {Collin}\ \emph {et~al.}(2016)\citenamefont {Collin},
  \citenamefont {Arg\"uelles}, \citenamefont {Conrad},\ and\ \citenamefont
  {Shaevitz}}]{Collin:2016aqd}%
  \BibitemOpen
  \bibfield  {author} {\bibinfo {author} {\bibfnamefont {G.~H.}\ \bibnamefont
  {Collin}}, \bibinfo {author} {\bibfnamefont {C.~A.}\ \bibnamefont
  {Arg\"uelles}}, \bibinfo {author} {\bibfnamefont {J.~M.}\ \bibnamefont
  {Conrad}}, \ and\ \bibinfo {author} {\bibfnamefont {M.~H.}\ \bibnamefont
  {Shaevitz}},\ }\href {\doibase 10.1103/PhysRevLett.117.221801} {\bibfield
  {journal} {\bibinfo  {journal} {Phys. Rev. Lett.}\ }\textbf {\bibinfo
  {volume} {117}},\ \bibinfo {pages} {221801} (\bibinfo {year} {2016})},\
  \Eprint {http://arxiv.org/abs/1607.00011} {arXiv:1607.00011 [hep-ph]}
  \BibitemShut {NoStop}%
\bibitem [{\citenamefont {Giunti}\ and\ \citenamefont
  {Laveder}(2011{\natexlab{a}})}]{Giunti:2011gz}%
  \BibitemOpen
  \bibfield  {author} {\bibinfo {author} {\bibfnamefont {C.}~\bibnamefont
  {Giunti}}\ and\ \bibinfo {author} {\bibfnamefont {M.}~\bibnamefont
  {Laveder}},\ }\href {\doibase 10.1103/PhysRevD.84.073008} {\bibfield
  {journal} {\bibinfo  {journal} {Phys. Rev. D}\ }\textbf {\bibinfo {volume}
  {84}},\ \bibinfo {pages} {073008} (\bibinfo {year} {2011}{\natexlab{a}})},\
  \Eprint {http://arxiv.org/abs/1107.1452} {arXiv:1107.1452 [hep-ph]}
  \BibitemShut {NoStop}%
\bibitem [{\citenamefont {Giunti}\ and\ \citenamefont
  {Laveder}(2011{\natexlab{b}})}]{Giunti:2011cp}%
  \BibitemOpen
  \bibfield  {author} {\bibinfo {author} {\bibfnamefont {C.}~\bibnamefont
  {Giunti}}\ and\ \bibinfo {author} {\bibfnamefont {M.}~\bibnamefont
  {Laveder}},\ }\href {\doibase 10.1016/j.physletb.2011.11.015} {\bibfield
  {journal} {\bibinfo  {journal} {Phys. Lett. B}\ }\textbf {\bibinfo {volume}
  {706}},\ \bibinfo {pages} {200} (\bibinfo {year} {2011}{\natexlab{b}})},\
  \Eprint {http://arxiv.org/abs/1111.1069} {arXiv:1111.1069 [hep-ph]}
  \BibitemShut {NoStop}%
\bibitem [{\citenamefont {Gariazzo}\ \emph {et~al.}(2017)\citenamefont
  {Gariazzo}, \citenamefont {Giunti}, \citenamefont {Laveder},\ and\
  \citenamefont {Li}}]{Gariazzo:2017fdh}%
  \BibitemOpen
  \bibfield  {author} {\bibinfo {author} {\bibfnamefont {S.}~\bibnamefont
  {Gariazzo}}, \bibinfo {author} {\bibfnamefont {C.}~\bibnamefont {Giunti}},
  \bibinfo {author} {\bibfnamefont {M.}~\bibnamefont {Laveder}}, \ and\
  \bibinfo {author} {\bibfnamefont {Y.~F.}\ \bibnamefont {Li}},\ }\href
  {\doibase 10.1007/JHEP06(2017)135} {\bibfield  {journal} {\bibinfo  {journal}
  {JHEP}\ }\textbf {\bibinfo {volume} {06}},\ \bibinfo {pages} {135} (\bibinfo
  {year} {2017})},\ \Eprint {http://arxiv.org/abs/1703.00860} {arXiv:1703.00860
  [hep-ph]} \BibitemShut {NoStop}%
\bibitem [{\citenamefont {B\"oser}\ \emph {et~al.}(2020)\citenamefont
  {B\"oser}, \citenamefont {Buck}, \citenamefont {Giunti}, \citenamefont
  {Lesgourgues}, \citenamefont {Ludhova}, \citenamefont {Mertens},
  \citenamefont {Schukraft},\ and\ \citenamefont {Wurm}}]{Boser:2019rta}%
  \BibitemOpen
  \bibfield  {author} {\bibinfo {author} {\bibfnamefont {S.}~\bibnamefont
  {B\"oser}}, \bibinfo {author} {\bibfnamefont {C.}~\bibnamefont {Buck}},
  \bibinfo {author} {\bibfnamefont {C.}~\bibnamefont {Giunti}}, \bibinfo
  {author} {\bibfnamefont {J.}~\bibnamefont {Lesgourgues}}, \bibinfo {author}
  {\bibfnamefont {L.}~\bibnamefont {Ludhova}}, \bibinfo {author} {\bibfnamefont
  {S.}~\bibnamefont {Mertens}}, \bibinfo {author} {\bibfnamefont
  {A.}~\bibnamefont {Schukraft}}, \ and\ \bibinfo {author} {\bibfnamefont
  {M.}~\bibnamefont {Wurm}},\ }\href {\doibase 10.1016/j.ppnp.2019.103736}
  {\bibfield  {journal} {\bibinfo  {journal} {Prog. Part. Nucl. Phys.}\
  }\textbf {\bibinfo {volume} {111}},\ \bibinfo {pages} {103736} (\bibinfo
  {year} {2020})},\ \Eprint {http://arxiv.org/abs/1906.01739} {arXiv:1906.01739
  [hep-ex]} \BibitemShut {NoStop}%
\bibitem [{\citenamefont {Kopp}\ \emph {et~al.}(2011)\citenamefont {Kopp},
  \citenamefont {Maltoni},\ and\ \citenamefont {Schwetz}}]{Kopp:2011qd}%
  \BibitemOpen
  \bibfield  {author} {\bibinfo {author} {\bibfnamefont {J.}~\bibnamefont
  {Kopp}}, \bibinfo {author} {\bibfnamefont {M.}~\bibnamefont {Maltoni}}, \
  and\ \bibinfo {author} {\bibfnamefont {T.}~\bibnamefont {Schwetz}},\ }\href
  {\doibase 10.1103/PhysRevLett.107.091801} {\bibfield  {journal} {\bibinfo
  {journal} {Phys. Rev. Lett.}\ }\textbf {\bibinfo {volume} {107}},\ \bibinfo
  {pages} {091801} (\bibinfo {year} {2011})},\ \Eprint
  {http://arxiv.org/abs/1103.4570} {arXiv:1103.4570 [hep-ph]} \BibitemShut
  {NoStop}%
\bibitem [{\citenamefont {Kopp}\ \emph {et~al.}(2013)\citenamefont {Kopp},
  \citenamefont {Machado}, \citenamefont {Maltoni},\ and\ \citenamefont
  {Schwetz}}]{Kopp:2013vaa}%
  \BibitemOpen
  \bibfield  {author} {\bibinfo {author} {\bibfnamefont {J.}~\bibnamefont
  {Kopp}}, \bibinfo {author} {\bibfnamefont {P.~A.~N.}\ \bibnamefont
  {Machado}}, \bibinfo {author} {\bibfnamefont {M.}~\bibnamefont {Maltoni}}, \
  and\ \bibinfo {author} {\bibfnamefont {T.}~\bibnamefont {Schwetz}},\ }\href
  {\doibase 10.1007/JHEP05(2013)050} {\bibfield  {journal} {\bibinfo  {journal}
  {JHEP}\ }\textbf {\bibinfo {volume} {05}},\ \bibinfo {pages} {050} (\bibinfo
  {year} {2013})},\ \Eprint {http://arxiv.org/abs/1303.3011} {arXiv:1303.3011
  [hep-ph]} \BibitemShut {NoStop}%
\bibitem [{\citenamefont {Dentler}\ \emph {et~al.}(2018)\citenamefont
  {Dentler}, \citenamefont {Hern\'andez-Cabezudo}, \citenamefont {Kopp},
  \citenamefont {Machado}, \citenamefont {Maltoni}, \citenamefont
  {Martinez-Soler},\ and\ \citenamefont {Schwetz}}]{Dentler:2018sju}%
  \BibitemOpen
  \bibfield  {author} {\bibinfo {author} {\bibfnamefont {M.}~\bibnamefont
  {Dentler}}, \bibinfo {author} {\bibfnamefont {A.}~\bibnamefont
  {Hern\'andez-Cabezudo}}, \bibinfo {author} {\bibfnamefont {J.}~\bibnamefont
  {Kopp}}, \bibinfo {author} {\bibfnamefont {P.~A.~N.}\ \bibnamefont
  {Machado}}, \bibinfo {author} {\bibfnamefont {M.}~\bibnamefont {Maltoni}},
  \bibinfo {author} {\bibfnamefont {I.}~\bibnamefont {Martinez-Soler}}, \ and\
  \bibinfo {author} {\bibfnamefont {T.}~\bibnamefont {Schwetz}},\ }\href
  {\doibase 10.1007/JHEP08(2018)010} {\bibfield  {journal} {\bibinfo  {journal}
  {JHEP}\ }\textbf {\bibinfo {volume} {08}},\ \bibinfo {pages} {010} (\bibinfo
  {year} {2018})},\ \Eprint {http://arxiv.org/abs/1803.10661} {arXiv:1803.10661
  [hep-ph]} \BibitemShut {NoStop}%
\bibitem [{\citenamefont {Abazajian}\ \emph {et~al.}(2012)\citenamefont
  {Abazajian} \emph {et~al.}}]{Abazajian:2012ys}%
  \BibitemOpen
  \bibfield  {author} {\bibinfo {author} {\bibfnamefont {K.~N.}\ \bibnamefont
  {Abazajian}} \emph {et~al.},\ }\href@noop {} {\  (\bibinfo {year} {2012})},\
  \Eprint {http://arxiv.org/abs/1204.5379} {arXiv:1204.5379 [hep-ph]}
  \BibitemShut {NoStop}%
\bibitem [{\citenamefont {Conrad}\ \emph {et~al.}(2013)\citenamefont {Conrad},
  \citenamefont {Ignarra}, \citenamefont {Karagiorgi}, \citenamefont
  {Shaevitz},\ and\ \citenamefont {Spitz}}]{Conrad:2012qt}%
  \BibitemOpen
  \bibfield  {author} {\bibinfo {author} {\bibfnamefont {J.~M.}\ \bibnamefont
  {Conrad}}, \bibinfo {author} {\bibfnamefont {C.~M.}\ \bibnamefont {Ignarra}},
  \bibinfo {author} {\bibfnamefont {G.}~\bibnamefont {Karagiorgi}}, \bibinfo
  {author} {\bibfnamefont {M.~H.}\ \bibnamefont {Shaevitz}}, \ and\ \bibinfo
  {author} {\bibfnamefont {J.}~\bibnamefont {Spitz}},\ }\href {\doibase
  10.1155/2013/163897} {\bibfield  {journal} {\bibinfo  {journal} {Adv. High
  Energy Phys.}\ }\textbf {\bibinfo {volume} {2013}},\ \bibinfo {pages}
  {163897} (\bibinfo {year} {2013})},\ \Eprint {http://arxiv.org/abs/1207.4765}
  {arXiv:1207.4765 [hep-ex]} \BibitemShut {NoStop}%
\bibitem [{\citenamefont {Diaz}\ \emph {et~al.}(2020)\citenamefont {Diaz},
  \citenamefont {Arg\"uelles}, \citenamefont {Collin}, \citenamefont {Conrad},\
  and\ \citenamefont {Shaevitz}}]{Diaz:2019fwt}%
  \BibitemOpen
  \bibfield  {author} {\bibinfo {author} {\bibfnamefont {A.}~\bibnamefont
  {Diaz}}, \bibinfo {author} {\bibfnamefont {C.~A.}\ \bibnamefont
  {Arg\"uelles}}, \bibinfo {author} {\bibfnamefont {G.~H.}\ \bibnamefont
  {Collin}}, \bibinfo {author} {\bibfnamefont {J.~M.}\ \bibnamefont {Conrad}},
  \ and\ \bibinfo {author} {\bibfnamefont {M.~H.}\ \bibnamefont {Shaevitz}},\
  }\href {\doibase 10.1016/j.physrep.2020.08.005} {\bibfield  {journal}
  {\bibinfo  {journal} {Phys. Rept.}\ }\textbf {\bibinfo {volume} {884}},\
  \bibinfo {pages} {1} (\bibinfo {year} {2020})},\ \Eprint
  {http://arxiv.org/abs/1906.00045} {arXiv:1906.00045 [hep-ex]} \BibitemShut
  {NoStop}%
\bibitem [{\citenamefont {Asaadi}\ \emph {et~al.}(2018)\citenamefont {Asaadi},
  \citenamefont {Church}, \citenamefont {Guenette}, \citenamefont {Jones},\
  and\ \citenamefont {Szelc}}]{Asaadi:2017bhx}%
  \BibitemOpen
  \bibfield  {author} {\bibinfo {author} {\bibfnamefont {J.}~\bibnamefont
  {Asaadi}}, \bibinfo {author} {\bibfnamefont {E.}~\bibnamefont {Church}},
  \bibinfo {author} {\bibfnamefont {R.}~\bibnamefont {Guenette}}, \bibinfo
  {author} {\bibfnamefont {B.~J.~P.}\ \bibnamefont {Jones}}, \ and\ \bibinfo
  {author} {\bibfnamefont {A.~M.}\ \bibnamefont {Szelc}},\ }\href {\doibase
  10.1103/PhysRevD.97.075021} {\bibfield  {journal} {\bibinfo  {journal} {Phys.
  Rev. D}\ }\textbf {\bibinfo {volume} {97}},\ \bibinfo {pages} {075021}
  (\bibinfo {year} {2018})},\ \Eprint {http://arxiv.org/abs/1712.08019}
  {arXiv:1712.08019 [hep-ph]} \BibitemShut {NoStop}%
\bibitem [{\citenamefont {Karagiorgi}\ \emph {et~al.}(2012)\citenamefont
  {Karagiorgi}, \citenamefont {Shaevitz},\ and\ \citenamefont
  {Conrad}}]{Karagiorgi:2012kw}%
  \BibitemOpen
  \bibfield  {author} {\bibinfo {author} {\bibfnamefont {G.}~\bibnamefont
  {Karagiorgi}}, \bibinfo {author} {\bibfnamefont {M.~H.}\ \bibnamefont
  {Shaevitz}}, \ and\ \bibinfo {author} {\bibfnamefont {J.~M.}\ \bibnamefont
  {Conrad}},\ }\href@noop {} {\  (\bibinfo {year} {2012})},\ \Eprint
  {http://arxiv.org/abs/1202.1024} {arXiv:1202.1024 [hep-ph]} \BibitemShut
  {NoStop}%
\bibitem [{\citenamefont {Pas}\ \emph {et~al.}(2005)\citenamefont {Pas},
  \citenamefont {Pakvasa},\ and\ \citenamefont {Weiler}}]{Pas:2005rb}%
  \BibitemOpen
  \bibfield  {author} {\bibinfo {author} {\bibfnamefont {H.}~\bibnamefont
  {Pas}}, \bibinfo {author} {\bibfnamefont {S.}~\bibnamefont {Pakvasa}}, \ and\
  \bibinfo {author} {\bibfnamefont {T.~J.}\ \bibnamefont {Weiler}},\ }\href
  {\doibase 10.1103/PhysRevD.72.095017} {\bibfield  {journal} {\bibinfo
  {journal} {Phys. Rev. D}\ }\textbf {\bibinfo {volume} {72}},\ \bibinfo
  {pages} {095017} (\bibinfo {year} {2005})},\ \Eprint
  {http://arxiv.org/abs/hep-ph/0504096} {arXiv:hep-ph/0504096} \BibitemShut
  {NoStop}%
\bibitem [{\citenamefont {D\"oring}\ \emph {et~al.}(2020)\citenamefont
  {D\"oring}, \citenamefont {P\"as}, \citenamefont {Sicking},\ and\
  \citenamefont {Weiler}}]{Doring:2018cob}%
  \BibitemOpen
  \bibfield  {author} {\bibinfo {author} {\bibfnamefont {D.}~\bibnamefont
  {D\"oring}}, \bibinfo {author} {\bibfnamefont {H.}~\bibnamefont {P\"as}},
  \bibinfo {author} {\bibfnamefont {P.}~\bibnamefont {Sicking}}, \ and\
  \bibinfo {author} {\bibfnamefont {T.~J.}\ \bibnamefont {Weiler}},\ }\href
  {\doibase 10.1140/epjc/s10052-020-08720-2} {\bibfield  {journal} {\bibinfo
  {journal} {Eur. Phys. J. C}\ }\textbf {\bibinfo {volume} {80}},\ \bibinfo
  {pages} {1202} (\bibinfo {year} {2020})},\ \Eprint
  {http://arxiv.org/abs/1808.07460} {arXiv:1808.07460 [hep-ph]} \BibitemShut
  {NoStop}%
\bibitem [{\citenamefont {Kostelecky}\ and\ \citenamefont
  {Mewes}(2004)}]{Kostelecky:2003cr}%
  \BibitemOpen
  \bibfield  {author} {\bibinfo {author} {\bibfnamefont {V.~A.}\ \bibnamefont
  {Kostelecky}}\ and\ \bibinfo {author} {\bibfnamefont {M.}~\bibnamefont
  {Mewes}},\ }\href {\doibase 10.1103/PhysRevD.69.016005} {\bibfield  {journal}
  {\bibinfo  {journal} {Phys. Rev. D}\ }\textbf {\bibinfo {volume} {69}},\
  \bibinfo {pages} {016005} (\bibinfo {year} {2004})},\ \Eprint
  {http://arxiv.org/abs/hep-ph/0309025} {arXiv:hep-ph/0309025} \BibitemShut
  {NoStop}%
\bibitem [{\citenamefont {Katori}\ \emph {et~al.}(2006)\citenamefont {Katori},
  \citenamefont {Kostelecky},\ and\ \citenamefont {Tayloe}}]{Katori:2006mz}%
  \BibitemOpen
  \bibfield  {author} {\bibinfo {author} {\bibfnamefont {T.}~\bibnamefont
  {Katori}}, \bibinfo {author} {\bibfnamefont {V.~A.}\ \bibnamefont
  {Kostelecky}}, \ and\ \bibinfo {author} {\bibfnamefont {R.}~\bibnamefont
  {Tayloe}},\ }\href {\doibase 10.1103/PhysRevD.74.105009} {\bibfield
  {journal} {\bibinfo  {journal} {Phys. Rev. D}\ }\textbf {\bibinfo {volume}
  {74}},\ \bibinfo {pages} {105009} (\bibinfo {year} {2006})},\ \Eprint
  {http://arxiv.org/abs/hep-ph/0606154} {arXiv:hep-ph/0606154} \BibitemShut
  {NoStop}%
\bibitem [{\citenamefont {Diaz}\ and\ \citenamefont
  {Kostelecky}(2011)}]{Diaz:2010ft}%
  \BibitemOpen
  \bibfield  {author} {\bibinfo {author} {\bibfnamefont {J.~S.}\ \bibnamefont
  {Diaz}}\ and\ \bibinfo {author} {\bibfnamefont {V.~A.}\ \bibnamefont
  {Kostelecky}},\ }\href {\doibase 10.1016/j.physletb.2011.04.049} {\bibfield
  {journal} {\bibinfo  {journal} {Phys. Lett. B}\ }\textbf {\bibinfo {volume}
  {700}},\ \bibinfo {pages} {25} (\bibinfo {year} {2011})},\ \Eprint
  {http://arxiv.org/abs/1012.5985} {arXiv:1012.5985 [hep-ph]} \BibitemShut
  {NoStop}%
\bibitem [{\citenamefont {Diaz}\ and\ \citenamefont
  {Kostelecky}(2012)}]{Diaz:2011ia}%
  \BibitemOpen
  \bibfield  {author} {\bibinfo {author} {\bibfnamefont {J.~S.}\ \bibnamefont
  {Diaz}}\ and\ \bibinfo {author} {\bibfnamefont {A.}~\bibnamefont
  {Kostelecky}},\ }\href {\doibase 10.1103/PhysRevD.85.016013} {\bibfield
  {journal} {\bibinfo  {journal} {Phys. Rev. D}\ }\textbf {\bibinfo {volume}
  {85}},\ \bibinfo {pages} {016013} (\bibinfo {year} {2012})},\ \Eprint
  {http://arxiv.org/abs/1108.1799} {arXiv:1108.1799 [hep-ph]} \BibitemShut
  {NoStop}%
\bibitem [{\citenamefont {Gninenko}(2009)}]{Gninenko:2009ks}%
  \BibitemOpen
  \bibfield  {author} {\bibinfo {author} {\bibfnamefont {S.~N.}\ \bibnamefont
  {Gninenko}},\ }\href {\doibase 10.1103/PhysRevLett.103.241802} {\bibfield
  {journal} {\bibinfo  {journal} {Phys. Rev. Lett.}\ }\textbf {\bibinfo
  {volume} {103}},\ \bibinfo {pages} {241802} (\bibinfo {year} {2009})},\
  \Eprint {http://arxiv.org/abs/0902.3802} {arXiv:0902.3802 [hep-ph]}
  \BibitemShut {NoStop}%
\bibitem [{\citenamefont {Gninenko}\ and\ \citenamefont
  {Gorbunov}(2010)}]{Gninenko:2009yf}%
  \BibitemOpen
  \bibfield  {author} {\bibinfo {author} {\bibfnamefont {S.~N.}\ \bibnamefont
  {Gninenko}}\ and\ \bibinfo {author} {\bibfnamefont {D.~S.}\ \bibnamefont
  {Gorbunov}},\ }\href {\doibase 10.1103/PhysRevD.81.075013} {\bibfield
  {journal} {\bibinfo  {journal} {Phys. Rev. D}\ }\textbf {\bibinfo {volume}
  {81}},\ \bibinfo {pages} {075013} (\bibinfo {year} {2010})},\ \Eprint
  {http://arxiv.org/abs/0907.4666} {arXiv:0907.4666 [hep-ph]} \BibitemShut
  {NoStop}%
\bibitem [{\citenamefont {Bai}\ \emph {et~al.}(2016)\citenamefont {Bai},
  \citenamefont {Lu}, \citenamefont {Lu}, \citenamefont {Salvado},\ and\
  \citenamefont {Stefanek}}]{Bai:2015ztj}%
  \BibitemOpen
  \bibfield  {author} {\bibinfo {author} {\bibfnamefont {Y.}~\bibnamefont
  {Bai}}, \bibinfo {author} {\bibfnamefont {R.}~\bibnamefont {Lu}}, \bibinfo
  {author} {\bibfnamefont {S.}~\bibnamefont {Lu}}, \bibinfo {author}
  {\bibfnamefont {J.}~\bibnamefont {Salvado}}, \ and\ \bibinfo {author}
  {\bibfnamefont {B.~A.}\ \bibnamefont {Stefanek}},\ }\href {\doibase
  10.1103/PhysRevD.93.073004} {\bibfield  {journal} {\bibinfo  {journal} {Phys.
  Rev. D}\ }\textbf {\bibinfo {volume} {93}},\ \bibinfo {pages} {073004}
  (\bibinfo {year} {2016})},\ \Eprint {http://arxiv.org/abs/1512.05357}
  {arXiv:1512.05357 [hep-ph]} \BibitemShut {NoStop}%
\bibitem [{\citenamefont {Moss}\ \emph {et~al.}(2018)\citenamefont {Moss},
  \citenamefont {Moulai}, \citenamefont {Arg\"uelles},\ and\ \citenamefont
  {Conrad}}]{Moss:2017pur}%
  \BibitemOpen
  \bibfield  {author} {\bibinfo {author} {\bibfnamefont {Z.}~\bibnamefont
  {Moss}}, \bibinfo {author} {\bibfnamefont {M.~H.}\ \bibnamefont {Moulai}},
  \bibinfo {author} {\bibfnamefont {C.~A.}\ \bibnamefont {Arg\"uelles}}, \ and\
  \bibinfo {author} {\bibfnamefont {J.~M.}\ \bibnamefont {Conrad}},\ }\href
  {\doibase 10.1103/PhysRevD.97.055017} {\bibfield  {journal} {\bibinfo
  {journal} {Phys. Rev. D}\ }\textbf {\bibinfo {volume} {97}},\ \bibinfo
  {pages} {055017} (\bibinfo {year} {2018})},\ \Eprint
  {http://arxiv.org/abs/1711.05921} {arXiv:1711.05921 [hep-ph]} \BibitemShut
  {NoStop}%
\bibitem [{\citenamefont {Bertuzzo}\ \emph {et~al.}(2018)\citenamefont
  {Bertuzzo}, \citenamefont {Jana}, \citenamefont {Machado},\ and\
  \citenamefont {Zukanovich~Funchal}}]{Bertuzzo:2018itn}%
  \BibitemOpen
  \bibfield  {author} {\bibinfo {author} {\bibfnamefont {E.}~\bibnamefont
  {Bertuzzo}}, \bibinfo {author} {\bibfnamefont {S.}~\bibnamefont {Jana}},
  \bibinfo {author} {\bibfnamefont {P.~A.~N.}\ \bibnamefont {Machado}}, \ and\
  \bibinfo {author} {\bibfnamefont {R.}~\bibnamefont {Zukanovich~Funchal}},\
  }\href {\doibase 10.1103/PhysRevLett.121.241801} {\bibfield  {journal}
  {\bibinfo  {journal} {Phys. Rev. Lett.}\ }\textbf {\bibinfo {volume} {121}},\
  \bibinfo {pages} {241801} (\bibinfo {year} {2018})},\ \Eprint
  {http://arxiv.org/abs/1807.09877} {arXiv:1807.09877 [hep-ph]} \BibitemShut
  {NoStop}%
\bibitem [{\citenamefont {Ballett}\ \emph {et~al.}(2019)\citenamefont
  {Ballett}, \citenamefont {Pascoli},\ and\ \citenamefont
  {Ross-Lonergan}}]{Ballett:2018ynz}%
  \BibitemOpen
  \bibfield  {author} {\bibinfo {author} {\bibfnamefont {P.}~\bibnamefont
  {Ballett}}, \bibinfo {author} {\bibfnamefont {S.}~\bibnamefont {Pascoli}}, \
  and\ \bibinfo {author} {\bibfnamefont {M.}~\bibnamefont {Ross-Lonergan}},\
  }\href {\doibase 10.1103/PhysRevD.99.071701} {\bibfield  {journal} {\bibinfo
  {journal} {Phys. Rev. D}\ }\textbf {\bibinfo {volume} {99}},\ \bibinfo
  {pages} {071701} (\bibinfo {year} {2019})},\ \Eprint
  {http://arxiv.org/abs/1808.02915} {arXiv:1808.02915 [hep-ph]} \BibitemShut
  {NoStop}%
\bibitem [{\citenamefont {Fischer}\ \emph {et~al.}(2020)\citenamefont
  {Fischer}, \citenamefont {Hern\'andez-Cabezudo},\ and\ \citenamefont
  {Schwetz}}]{Fischer:2019fbw}%
  \BibitemOpen
  \bibfield  {author} {\bibinfo {author} {\bibfnamefont {O.}~\bibnamefont
  {Fischer}}, \bibinfo {author} {\bibfnamefont {A.}~\bibnamefont
  {Hern\'andez-Cabezudo}}, \ and\ \bibinfo {author} {\bibfnamefont
  {T.}~\bibnamefont {Schwetz}},\ }\href {\doibase 10.1103/PhysRevD.101.075045}
  {\bibfield  {journal} {\bibinfo  {journal} {Phys. Rev. D}\ }\textbf {\bibinfo
  {volume} {101}},\ \bibinfo {pages} {075045} (\bibinfo {year} {2020})},\
  \Eprint {http://arxiv.org/abs/1909.09561} {arXiv:1909.09561 [hep-ph]}
  \BibitemShut {NoStop}%
\bibitem [{\citenamefont {Moulai}\ \emph {et~al.}(2020)\citenamefont {Moulai},
  \citenamefont {Arg\"uelles}, \citenamefont {Collin}, \citenamefont {Conrad},
  \citenamefont {Diaz},\ and\ \citenamefont {Shaevitz}}]{Moulai:2019gpi}%
  \BibitemOpen
  \bibfield  {author} {\bibinfo {author} {\bibfnamefont {M.~H.}\ \bibnamefont
  {Moulai}}, \bibinfo {author} {\bibfnamefont {C.~A.}\ \bibnamefont
  {Arg\"uelles}}, \bibinfo {author} {\bibfnamefont {G.~H.}\ \bibnamefont
  {Collin}}, \bibinfo {author} {\bibfnamefont {J.~M.}\ \bibnamefont {Conrad}},
  \bibinfo {author} {\bibfnamefont {A.}~\bibnamefont {Diaz}}, \ and\ \bibinfo
  {author} {\bibfnamefont {M.~H.}\ \bibnamefont {Shaevitz}},\ }\href {\doibase
  10.1103/PhysRevD.101.055020} {\bibfield  {journal} {\bibinfo  {journal}
  {Phys. Rev. D}\ }\textbf {\bibinfo {volume} {101}},\ \bibinfo {pages}
  {055020} (\bibinfo {year} {2020})},\ \Eprint
  {http://arxiv.org/abs/1910.13456} {arXiv:1910.13456 [hep-ph]} \BibitemShut
  {NoStop}%
\bibitem [{\citenamefont {Dentler}\ \emph {et~al.}(2020)\citenamefont
  {Dentler}, \citenamefont {Esteban}, \citenamefont {Kopp},\ and\ \citenamefont
  {Machado}}]{Dentler:2019dhz}%
  \BibitemOpen
  \bibfield  {author} {\bibinfo {author} {\bibfnamefont {M.}~\bibnamefont
  {Dentler}}, \bibinfo {author} {\bibfnamefont {I.}~\bibnamefont {Esteban}},
  \bibinfo {author} {\bibfnamefont {J.}~\bibnamefont {Kopp}}, \ and\ \bibinfo
  {author} {\bibfnamefont {P.}~\bibnamefont {Machado}},\ }\href {\doibase
  10.1103/PhysRevD.101.115013} {\bibfield  {journal} {\bibinfo  {journal}
  {Phys. Rev. D}\ }\textbf {\bibinfo {volume} {101}},\ \bibinfo {pages}
  {115013} (\bibinfo {year} {2020})},\ \Eprint
  {http://arxiv.org/abs/1911.01427} {arXiv:1911.01427 [hep-ph]} \BibitemShut
  {NoStop}%
\bibitem [{\citenamefont {de~Gouv\^ea}\ \emph {et~al.}(2020)\citenamefont
  {de~Gouv\^ea}, \citenamefont {Peres}, \citenamefont {Prakash},\ and\
  \citenamefont {Stenico}}]{deGouvea:2019qre}%
  \BibitemOpen
  \bibfield  {author} {\bibinfo {author} {\bibfnamefont {A.}~\bibnamefont
  {de~Gouv\^ea}}, \bibinfo {author} {\bibfnamefont {O.~L.~G.}\ \bibnamefont
  {Peres}}, \bibinfo {author} {\bibfnamefont {S.}~\bibnamefont {Prakash}}, \
  and\ \bibinfo {author} {\bibfnamefont {G.~V.}\ \bibnamefont {Stenico}},\
  }\href {\doibase 10.1007/JHEP07(2020)141} {\bibfield  {journal} {\bibinfo
  {journal} {JHEP}\ }\textbf {\bibinfo {volume} {07}},\ \bibinfo {pages} {141}
  (\bibinfo {year} {2020})},\ \Eprint {http://arxiv.org/abs/1911.01447}
  {arXiv:1911.01447 [hep-ph]} \BibitemShut {NoStop}%
\bibitem [{\citenamefont {Datta}\ \emph {et~al.}(2020)\citenamefont {Datta},
  \citenamefont {Kamali},\ and\ \citenamefont {Marfatia}}]{Datta:2020auq}%
  \BibitemOpen
  \bibfield  {author} {\bibinfo {author} {\bibfnamefont {A.}~\bibnamefont
  {Datta}}, \bibinfo {author} {\bibfnamefont {S.}~\bibnamefont {Kamali}}, \
  and\ \bibinfo {author} {\bibfnamefont {D.}~\bibnamefont {Marfatia}},\ }\href
  {\doibase 10.1016/j.physletb.2020.135579} {\bibfield  {journal} {\bibinfo
  {journal} {Phys. Lett. B}\ }\textbf {\bibinfo {volume} {807}},\ \bibinfo
  {pages} {135579} (\bibinfo {year} {2020})},\ \Eprint
  {http://arxiv.org/abs/2005.08920} {arXiv:2005.08920 [hep-ph]} \BibitemShut
  {NoStop}%
\bibitem [{\citenamefont {Dutta}\ \emph
  {et~al.}(2020{\natexlab{a}})\citenamefont {Dutta}, \citenamefont {Ghosh},\
  and\ \citenamefont {Li}}]{Dutta:2020scq}%
  \BibitemOpen
  \bibfield  {author} {\bibinfo {author} {\bibfnamefont {B.}~\bibnamefont
  {Dutta}}, \bibinfo {author} {\bibfnamefont {S.}~\bibnamefont {Ghosh}}, \ and\
  \bibinfo {author} {\bibfnamefont {T.}~\bibnamefont {Li}},\ }\href {\doibase
  10.1103/PhysRevD.102.055017} {\bibfield  {journal} {\bibinfo  {journal}
  {Phys. Rev. D}\ }\textbf {\bibinfo {volume} {102}},\ \bibinfo {pages}
  {055017} (\bibinfo {year} {2020}{\natexlab{a}})},\ \Eprint
  {http://arxiv.org/abs/2006.01319} {arXiv:2006.01319 [hep-ph]} \BibitemShut
  {NoStop}%
\bibitem [{\citenamefont {Abdallah}\ \emph {et~al.}(2020)\citenamefont
  {Abdallah}, \citenamefont {Gandhi},\ and\ \citenamefont
  {Roy}}]{Abdallah:2020biq}%
  \BibitemOpen
  \bibfield  {author} {\bibinfo {author} {\bibfnamefont {W.}~\bibnamefont
  {Abdallah}}, \bibinfo {author} {\bibfnamefont {R.}~\bibnamefont {Gandhi}}, \
  and\ \bibinfo {author} {\bibfnamefont {S.}~\bibnamefont {Roy}},\ }\href
  {\doibase 10.1007/JHEP12(2020)188} {\bibfield  {journal} {\bibinfo  {journal}
  {JHEP}\ }\textbf {\bibinfo {volume} {12}},\ \bibinfo {pages} {188} (\bibinfo
  {year} {2020})},\ \Eprint {http://arxiv.org/abs/2006.01948} {arXiv:2006.01948
  [hep-ph]} \BibitemShut {NoStop}%
\bibitem [{\citenamefont {Abdullahi}\ \emph {et~al.}(2021)\citenamefont
  {Abdullahi}, \citenamefont {Hostert},\ and\ \citenamefont
  {Pascoli}}]{Abdullahi:2020nyr}%
  \BibitemOpen
  \bibfield  {author} {\bibinfo {author} {\bibfnamefont {A.}~\bibnamefont
  {Abdullahi}}, \bibinfo {author} {\bibfnamefont {M.}~\bibnamefont {Hostert}},
  \ and\ \bibinfo {author} {\bibfnamefont {S.}~\bibnamefont {Pascoli}},\ }\href
  {\doibase 10.1016/j.physletb.2021.136531} {\bibfield  {journal} {\bibinfo
  {journal} {Phys. Lett. B}\ }\textbf {\bibinfo {volume} {820}},\ \bibinfo
  {pages} {136531} (\bibinfo {year} {2021})},\ \Eprint
  {http://arxiv.org/abs/2007.11813} {arXiv:2007.11813 [hep-ph]} \BibitemShut
  {NoStop}%
\bibitem [{\citenamefont {Liao}\ and\ \citenamefont
  {Marfatia}(2016)}]{Liao:2016reh}%
  \BibitemOpen
  \bibfield  {author} {\bibinfo {author} {\bibfnamefont {J.}~\bibnamefont
  {Liao}}\ and\ \bibinfo {author} {\bibfnamefont {D.}~\bibnamefont
  {Marfatia}},\ }\href {\doibase 10.1103/PhysRevLett.117.071802} {\bibfield
  {journal} {\bibinfo  {journal} {Phys. Rev. Lett.}\ }\textbf {\bibinfo
  {volume} {117}},\ \bibinfo {pages} {071802} (\bibinfo {year} {2016})},\
  \Eprint {http://arxiv.org/abs/1602.08766} {arXiv:1602.08766 [hep-ph]}
  \BibitemShut {NoStop}%
\bibitem [{\citenamefont {Carena}\ \emph {et~al.}(2017)\citenamefont {Carena},
  \citenamefont {Li}, \citenamefont {Machado}, \citenamefont {Machado},\ and\
  \citenamefont {Wagner}}]{Carena:2017qhd}%
  \BibitemOpen
  \bibfield  {author} {\bibinfo {author} {\bibfnamefont {M.}~\bibnamefont
  {Carena}}, \bibinfo {author} {\bibfnamefont {Y.-Y.}\ \bibnamefont {Li}},
  \bibinfo {author} {\bibfnamefont {C.~S.}\ \bibnamefont {Machado}}, \bibinfo
  {author} {\bibfnamefont {P.~A.~N.}\ \bibnamefont {Machado}}, \ and\ \bibinfo
  {author} {\bibfnamefont {C.~E.~M.}\ \bibnamefont {Wagner}},\ }\href {\doibase
  10.1103/PhysRevD.96.095014} {\bibfield  {journal} {\bibinfo  {journal} {Phys.
  Rev. D}\ }\textbf {\bibinfo {volume} {96}},\ \bibinfo {pages} {095014}
  (\bibinfo {year} {2017})},\ \Eprint {http://arxiv.org/abs/1708.09548}
  {arXiv:1708.09548 [hep-ph]} \BibitemShut {NoStop}%
\bibitem [{\citenamefont {Abdallah}\ \emph {et~al.}(2021)\citenamefont
  {Abdallah}, \citenamefont {Gandhi},\ and\ \citenamefont
  {Roy}}]{Abdallah:2020vgg}%
  \BibitemOpen
  \bibfield  {author} {\bibinfo {author} {\bibfnamefont {W.}~\bibnamefont
  {Abdallah}}, \bibinfo {author} {\bibfnamefont {R.}~\bibnamefont {Gandhi}}, \
  and\ \bibinfo {author} {\bibfnamefont {S.}~\bibnamefont {Roy}},\ }\href
  {\doibase 10.1103/PhysRevD.104.055028} {\bibfield  {journal} {\bibinfo
  {journal} {Phys. Rev. D}\ }\textbf {\bibinfo {volume} {104}},\ \bibinfo
  {pages} {055028} (\bibinfo {year} {2021})},\ \Eprint
  {http://arxiv.org/abs/2010.06159} {arXiv:2010.06159 [hep-ph]} \BibitemShut
  {NoStop}%
\bibitem [{\citenamefont {Aguilar-Arevalo}\ \emph
  {et~al.}(2018{\natexlab{b}})\citenamefont {Aguilar-Arevalo} \emph
  {et~al.}}]{MiniBooNEDM:2018cxm}%
  \BibitemOpen
  \bibfield  {author} {\bibinfo {author} {\bibfnamefont {A.~A.}\ \bibnamefont
  {Aguilar-Arevalo}} \emph {et~al.} (\bibinfo {collaboration} {MiniBooNE DM}),\
  }\href {\doibase 10.1103/PhysRevD.98.112004} {\bibfield  {journal} {\bibinfo
  {journal} {Phys. Rev. D}\ }\textbf {\bibinfo {volume} {98}},\ \bibinfo
  {pages} {112004} (\bibinfo {year} {2018}{\natexlab{b}})},\ \Eprint
  {http://arxiv.org/abs/1807.06137} {arXiv:1807.06137 [hep-ex]} \BibitemShut
  {NoStop}%
\bibitem [{\citenamefont {Jordan}\ \emph {et~al.}(2019)\citenamefont {Jordan},
  \citenamefont {Kahn}, \citenamefont {Krnjaic}, \citenamefont {Moschella},\
  and\ \citenamefont {Spitz}}]{Jordan:2018qiy}%
  \BibitemOpen
  \bibfield  {author} {\bibinfo {author} {\bibfnamefont {J.~R.}\ \bibnamefont
  {Jordan}}, \bibinfo {author} {\bibfnamefont {Y.}~\bibnamefont {Kahn}},
  \bibinfo {author} {\bibfnamefont {G.}~\bibnamefont {Krnjaic}}, \bibinfo
  {author} {\bibfnamefont {M.}~\bibnamefont {Moschella}}, \ and\ \bibinfo
  {author} {\bibfnamefont {J.}~\bibnamefont {Spitz}},\ }\href {\doibase
  10.1103/PhysRevLett.122.081801} {\bibfield  {journal} {\bibinfo  {journal}
  {Phys. Rev. Lett.}\ }\textbf {\bibinfo {volume} {122}},\ \bibinfo {pages}
  {081801} (\bibinfo {year} {2019})},\ \Eprint
  {http://arxiv.org/abs/1810.07185} {arXiv:1810.07185 [hep-ph]} \BibitemShut
  {NoStop}%
\bibitem [{\citenamefont {Arg\"uelles}\ \emph {et~al.}(2019)\citenamefont
  {Arg\"uelles}, \citenamefont {Hostert},\ and\ \citenamefont
  {Tsai}}]{Arguelles:2018mtc}%
  \BibitemOpen
  \bibfield  {author} {\bibinfo {author} {\bibfnamefont {C.~A.}\ \bibnamefont
  {Arg\"uelles}}, \bibinfo {author} {\bibfnamefont {M.}~\bibnamefont
  {Hostert}}, \ and\ \bibinfo {author} {\bibfnamefont {Y.-D.}\ \bibnamefont
  {Tsai}},\ }\href {\doibase 10.1103/PhysRevLett.123.261801} {\bibfield
  {journal} {\bibinfo  {journal} {Phys. Rev. Lett.}\ }\textbf {\bibinfo
  {volume} {123}},\ \bibinfo {pages} {261801} (\bibinfo {year} {2019})},\
  \Eprint {http://arxiv.org/abs/1812.08768} {arXiv:1812.08768 [hep-ph]}
  \BibitemShut {NoStop}%
\bibitem [{\citenamefont {Brdar}\ \emph {et~al.}(2021)\citenamefont {Brdar},
  \citenamefont {Fischer},\ and\ \citenamefont {Smirnov}}]{Brdar:2020tle}%
  \BibitemOpen
  \bibfield  {author} {\bibinfo {author} {\bibfnamefont {V.}~\bibnamefont
  {Brdar}}, \bibinfo {author} {\bibfnamefont {O.}~\bibnamefont {Fischer}}, \
  and\ \bibinfo {author} {\bibfnamefont {A.~Y.}\ \bibnamefont {Smirnov}},\
  }\href {\doibase 10.1103/PhysRevD.103.075008} {\bibfield  {journal} {\bibinfo
   {journal} {Phys. Rev. D}\ }\textbf {\bibinfo {volume} {103}},\ \bibinfo
  {pages} {075008} (\bibinfo {year} {2021})},\ \Eprint
  {http://arxiv.org/abs/2007.14411} {arXiv:2007.14411 [hep-ph]} \BibitemShut
  {NoStop}%
\bibitem [{\citenamefont {Barger}\ \emph {et~al.}(2012)\citenamefont {Barger},
  \citenamefont {Chiang}, \citenamefont {Keung},\ and\ \citenamefont
  {Marfatia}}]{Barger:2011mt}%
  \BibitemOpen
  \bibfield  {author} {\bibinfo {author} {\bibfnamefont {V.}~\bibnamefont
  {Barger}}, \bibinfo {author} {\bibfnamefont {C.-W.}\ \bibnamefont {Chiang}},
  \bibinfo {author} {\bibfnamefont {W.-Y.}\ \bibnamefont {Keung}}, \ and\
  \bibinfo {author} {\bibfnamefont {D.}~\bibnamefont {Marfatia}},\ }\href
  {\doibase 10.1103/PhysRevLett.108.081802} {\bibfield  {journal} {\bibinfo
  {journal} {Phys. Rev. Lett.}\ }\textbf {\bibinfo {volume} {108}},\ \bibinfo
  {pages} {081802} (\bibinfo {year} {2012})},\ \Eprint
  {http://arxiv.org/abs/1109.6652} {arXiv:1109.6652 [hep-ph]} \BibitemShut
  {NoStop}%
\bibitem [{\citenamefont {Carlson}\ and\ \citenamefont
  {Rislow}(2012)}]{Carlson:2012pc}%
  \BibitemOpen
  \bibfield  {author} {\bibinfo {author} {\bibfnamefont {C.~E.}\ \bibnamefont
  {Carlson}}\ and\ \bibinfo {author} {\bibfnamefont {B.~C.}\ \bibnamefont
  {Rislow}},\ }\href {\doibase 10.1103/PhysRevD.86.035013} {\bibfield
  {journal} {\bibinfo  {journal} {Phys. Rev. D}\ }\textbf {\bibinfo {volume}
  {86}},\ \bibinfo {pages} {035013} (\bibinfo {year} {2012})},\ \Eprint
  {http://arxiv.org/abs/1206.3587} {arXiv:1206.3587 [hep-ph]} \BibitemShut
  {NoStop}%
\bibitem [{\citenamefont {Laha}\ \emph {et~al.}(2014)\citenamefont {Laha},
  \citenamefont {Dasgupta},\ and\ \citenamefont {Beacom}}]{Laha:2013xua}%
  \BibitemOpen
  \bibfield  {author} {\bibinfo {author} {\bibfnamefont {R.}~\bibnamefont
  {Laha}}, \bibinfo {author} {\bibfnamefont {B.}~\bibnamefont {Dasgupta}}, \
  and\ \bibinfo {author} {\bibfnamefont {J.~F.}\ \bibnamefont {Beacom}},\
  }\href {\doibase 10.1103/PhysRevD.89.093025} {\bibfield  {journal} {\bibinfo
  {journal} {Phys. Rev. D}\ }\textbf {\bibinfo {volume} {89}},\ \bibinfo
  {pages} {093025} (\bibinfo {year} {2014})},\ \Eprint
  {http://arxiv.org/abs/1304.3460} {arXiv:1304.3460 [hep-ph]} \BibitemShut
  {NoStop}%
\bibitem [{\citenamefont {Bakhti}\ and\ \citenamefont
  {Farzan}(2017)}]{Bakhti:2017jhm}%
  \BibitemOpen
  \bibfield  {author} {\bibinfo {author} {\bibfnamefont {P.}~\bibnamefont
  {Bakhti}}\ and\ \bibinfo {author} {\bibfnamefont {Y.}~\bibnamefont
  {Farzan}},\ }\href {\doibase 10.1103/PhysRevD.95.095008} {\bibfield
  {journal} {\bibinfo  {journal} {Phys. Rev. D}\ }\textbf {\bibinfo {volume}
  {95}},\ \bibinfo {pages} {095008} (\bibinfo {year} {2017})},\ \Eprint
  {http://arxiv.org/abs/1702.04187} {arXiv:1702.04187 [hep-ph]} \BibitemShut
  {NoStop}%
\bibitem [{\citenamefont {Krnjaic}\ \emph {et~al.}(2020)\citenamefont
  {Krnjaic}, \citenamefont {Marques-Tavares}, \citenamefont {Redigolo},\ and\
  \citenamefont {Tobioka}}]{Krnjaic:2019rsv}%
  \BibitemOpen
  \bibfield  {author} {\bibinfo {author} {\bibfnamefont {G.}~\bibnamefont
  {Krnjaic}}, \bibinfo {author} {\bibfnamefont {G.}~\bibnamefont
  {Marques-Tavares}}, \bibinfo {author} {\bibfnamefont {D.}~\bibnamefont
  {Redigolo}}, \ and\ \bibinfo {author} {\bibfnamefont {K.}~\bibnamefont
  {Tobioka}},\ }\href {\doibase 10.1103/PhysRevLett.124.041802} {\bibfield
  {journal} {\bibinfo  {journal} {Phys. Rev. Lett.}\ }\textbf {\bibinfo
  {volume} {124}},\ \bibinfo {pages} {041802} (\bibinfo {year} {2020})},\
  \Eprint {http://arxiv.org/abs/1902.07715} {arXiv:1902.07715 [hep-ph]}
  \BibitemShut {NoStop}%
\bibitem [{\citenamefont {Zyla}\ \emph {et~al.}(2020)\citenamefont {Zyla} \emph
  {et~al.}}]{ParticleDataGroup:2020ssz}%
  \BibitemOpen
  \bibfield  {author} {\bibinfo {author} {\bibfnamefont {P.~A.}\ \bibnamefont
  {Zyla}} \emph {et~al.} (\bibinfo {collaboration} {Particle Data Group}),\
  }\href {\doibase 10.1093/ptep/ptaa104} {\bibfield  {journal} {\bibinfo
  {journal} {PTEP}\ }\textbf {\bibinfo {volume} {2020}},\ \bibinfo {pages}
  {083C01} (\bibinfo {year} {2020})}\BibitemShut {NoStop}%
\bibitem [{\citenamefont {Dutta}\ \emph
  {et~al.}(2020{\natexlab{b}})\citenamefont {Dutta}, \citenamefont {Kim},
  \citenamefont {Liao}, \citenamefont {Park}, \citenamefont {Shin},\ and\
  \citenamefont {Strigari}}]{Dutta:2019nbn}%
  \BibitemOpen
  \bibfield  {author} {\bibinfo {author} {\bibfnamefont {B.}~\bibnamefont
  {Dutta}}, \bibinfo {author} {\bibfnamefont {D.}~\bibnamefont {Kim}}, \bibinfo
  {author} {\bibfnamefont {S.}~\bibnamefont {Liao}}, \bibinfo {author}
  {\bibfnamefont {J.-C.}\ \bibnamefont {Park}}, \bibinfo {author}
  {\bibfnamefont {S.}~\bibnamefont {Shin}}, \ and\ \bibinfo {author}
  {\bibfnamefont {L.~E.}\ \bibnamefont {Strigari}},\ }\href {\doibase
  10.1103/PhysRevLett.124.121802} {\bibfield  {journal} {\bibinfo  {journal}
  {Phys. Rev. Lett.}\ }\textbf {\bibinfo {volume} {124}},\ \bibinfo {pages}
  {121802} (\bibinfo {year} {2020}{\natexlab{b}})},\ \Eprint
  {http://arxiv.org/abs/1906.10745} {arXiv:1906.10745 [hep-ph]} \BibitemShut
  {NoStop}%
\bibitem [{\citenamefont {Dutta}\ \emph
  {et~al.}(2020{\natexlab{c}})\citenamefont {Dutta}, \citenamefont {Kim},
  \citenamefont {Liao}, \citenamefont {Park}, \citenamefont {Shin},
  \citenamefont {Strigari},\ and\ \citenamefont {Thompson}}]{Dutta:2020vop}%
  \BibitemOpen
  \bibfield  {author} {\bibinfo {author} {\bibfnamefont {B.}~\bibnamefont
  {Dutta}}, \bibinfo {author} {\bibfnamefont {D.}~\bibnamefont {Kim}}, \bibinfo
  {author} {\bibfnamefont {S.}~\bibnamefont {Liao}}, \bibinfo {author}
  {\bibfnamefont {J.-C.}\ \bibnamefont {Park}}, \bibinfo {author}
  {\bibfnamefont {S.}~\bibnamefont {Shin}}, \bibinfo {author} {\bibfnamefont
  {L.~E.}\ \bibnamefont {Strigari}}, \ and\ \bibinfo {author} {\bibfnamefont
  {A.}~\bibnamefont {Thompson}},\ }\href@noop {} {\  (\bibinfo {year}
  {2020}{\natexlab{c}})},\ \Eprint {http://arxiv.org/abs/2006.09386}
  {arXiv:2006.09386 [hep-ph]} \BibitemShut {NoStop}%
\bibitem [{\citenamefont {Tucker-Smith}\ and\ \citenamefont
  {Weiner}(2001)}]{Tucker-Smith:2001myb}%
  \BibitemOpen
  \bibfield  {author} {\bibinfo {author} {\bibfnamefont {D.}~\bibnamefont
  {Tucker-Smith}}\ and\ \bibinfo {author} {\bibfnamefont {N.}~\bibnamefont
  {Weiner}},\ }\href {\doibase 10.1103/PhysRevD.64.043502} {\bibfield
  {journal} {\bibinfo  {journal} {Phys. Rev. D}\ }\textbf {\bibinfo {volume}
  {64}},\ \bibinfo {pages} {043502} (\bibinfo {year} {2001})},\ \Eprint
  {http://arxiv.org/abs/hep-ph/0101138} {arXiv:hep-ph/0101138} \BibitemShut
  {NoStop}%
\bibitem [{\citenamefont {Izaguirre}\ \emph {et~al.}(2014)\citenamefont
  {Izaguirre}, \citenamefont {Krnjaic}, \citenamefont {Schuster},\ and\
  \citenamefont {Toro}}]{Izaguirre:2014dua}%
  \BibitemOpen
  \bibfield  {author} {\bibinfo {author} {\bibfnamefont {E.}~\bibnamefont
  {Izaguirre}}, \bibinfo {author} {\bibfnamefont {G.}~\bibnamefont {Krnjaic}},
  \bibinfo {author} {\bibfnamefont {P.}~\bibnamefont {Schuster}}, \ and\
  \bibinfo {author} {\bibfnamefont {N.}~\bibnamefont {Toro}},\ }\href {\doibase
  10.1103/PhysRevD.90.014052} {\bibfield  {journal} {\bibinfo  {journal} {Phys.
  Rev. D}\ }\textbf {\bibinfo {volume} {90}},\ \bibinfo {pages} {014052}
  (\bibinfo {year} {2014})},\ \Eprint {http://arxiv.org/abs/1403.6826}
  {arXiv:1403.6826 [hep-ph]} \BibitemShut {NoStop}%
\bibitem [{\citenamefont {Giudice}\ \emph {et~al.}(2018)\citenamefont
  {Giudice}, \citenamefont {Kim}, \citenamefont {Park},\ and\ \citenamefont
  {Shin}}]{Giudice:2017zke}%
  \BibitemOpen
  \bibfield  {author} {\bibinfo {author} {\bibfnamefont {G.~F.}\ \bibnamefont
  {Giudice}}, \bibinfo {author} {\bibfnamefont {D.}~\bibnamefont {Kim}},
  \bibinfo {author} {\bibfnamefont {J.-C.}\ \bibnamefont {Park}}, \ and\
  \bibinfo {author} {\bibfnamefont {S.}~\bibnamefont {Shin}},\ }\href {\doibase
  10.1016/j.physletb.2018.03.043} {\bibfield  {journal} {\bibinfo  {journal}
  {Phys. Lett. B}\ }\textbf {\bibinfo {volume} {780}},\ \bibinfo {pages} {543}
  (\bibinfo {year} {2018})},\ \Eprint {http://arxiv.org/abs/1712.07126}
  {arXiv:1712.07126 [hep-ph]} \BibitemShut {NoStop}%
\bibitem [{\citenamefont {Dutta}\ \emph {et~al.}(2019)\citenamefont {Dutta},
  \citenamefont {Ghosh},\ and\ \citenamefont {Kumar}}]{Dutta:2019fxn}%
  \BibitemOpen
  \bibfield  {author} {\bibinfo {author} {\bibfnamefont {B.}~\bibnamefont
  {Dutta}}, \bibinfo {author} {\bibfnamefont {S.}~\bibnamefont {Ghosh}}, \ and\
  \bibinfo {author} {\bibfnamefont {J.}~\bibnamefont {Kumar}},\ }\href
  {\doibase 10.1103/PhysRevD.100.075028} {\bibfield  {journal} {\bibinfo
  {journal} {Phys. Rev. D}\ }\textbf {\bibinfo {volume} {100}},\ \bibinfo
  {pages} {075028} (\bibinfo {year} {2019})},\ \Eprint
  {http://arxiv.org/abs/1905.02692} {arXiv:1905.02692 [hep-ph]} \BibitemShut
  {NoStop}%
\bibitem [{\citenamefont {Dutta}\ \emph
  {et~al.}(2020{\natexlab{d}})\citenamefont {Dutta}, \citenamefont {Ghosh},\
  and\ \citenamefont {Kumar}}]{Dutta:2020enk}%
  \BibitemOpen
  \bibfield  {author} {\bibinfo {author} {\bibfnamefont {B.}~\bibnamefont
  {Dutta}}, \bibinfo {author} {\bibfnamefont {S.}~\bibnamefont {Ghosh}}, \ and\
  \bibinfo {author} {\bibfnamefont {J.}~\bibnamefont {Kumar}},\ }\href
  {\doibase 10.1103/PhysRevD.102.075041} {\bibfield  {journal} {\bibinfo
  {journal} {Phys. Rev. D}\ }\textbf {\bibinfo {volume} {102}},\ \bibinfo
  {pages} {075041} (\bibinfo {year} {2020}{\natexlab{d}})},\ \Eprint
  {http://arxiv.org/abs/2007.16191} {arXiv:2007.16191 [hep-ph]} \BibitemShut
  {NoStop}%
\bibitem [{\citenamefont {Kim}\ \emph {et~al.}(2017)\citenamefont {Kim},
  \citenamefont {Park},\ and\ \citenamefont {Shin}}]{Kim:2016zjx}%
  \BibitemOpen
  \bibfield  {author} {\bibinfo {author} {\bibfnamefont {D.}~\bibnamefont
  {Kim}}, \bibinfo {author} {\bibfnamefont {J.-C.}\ \bibnamefont {Park}}, \
  and\ \bibinfo {author} {\bibfnamefont {S.}~\bibnamefont {Shin}},\ }\href
  {\doibase 10.1103/PhysRevLett.119.161801} {\bibfield  {journal} {\bibinfo
  {journal} {Phys. Rev. Lett.}\ }\textbf {\bibinfo {volume} {119}},\ \bibinfo
  {pages} {161801} (\bibinfo {year} {2017})},\ \Eprint
  {http://arxiv.org/abs/1612.06867} {arXiv:1612.06867 [hep-ph]} \BibitemShut
  {NoStop}%
\bibitem [{\citenamefont {Kim}\ \emph {et~al.}(2020)\citenamefont {Kim},
  \citenamefont {Machado}, \citenamefont {Park},\ and\ \citenamefont
  {Shin}}]{Kim:2020ipj}%
  \BibitemOpen
  \bibfield  {author} {\bibinfo {author} {\bibfnamefont {D.}~\bibnamefont
  {Kim}}, \bibinfo {author} {\bibfnamefont {P.~A.~N.}\ \bibnamefont {Machado}},
  \bibinfo {author} {\bibfnamefont {J.-C.}\ \bibnamefont {Park}}, \ and\
  \bibinfo {author} {\bibfnamefont {S.}~\bibnamefont {Shin}},\ }\href {\doibase
  10.1007/JHEP07(2020)057} {\bibfield  {journal} {\bibinfo  {journal} {JHEP}\
  }\textbf {\bibinfo {volume} {07}},\ \bibinfo {pages} {057} (\bibinfo {year}
  {2020})},\ \Eprint {http://arxiv.org/abs/2003.07369} {arXiv:2003.07369
  [hep-ph]} \BibitemShut {NoStop}%
\bibitem [{\citenamefont {Kaneta}\ \emph
  {et~al.}(2017{\natexlab{a}})\citenamefont {Kaneta}, \citenamefont {Lee},\
  and\ \citenamefont {Yun}}]{Kaneta:2016wvf}%
  \BibitemOpen
  \bibfield  {author} {\bibinfo {author} {\bibfnamefont {K.}~\bibnamefont
  {Kaneta}}, \bibinfo {author} {\bibfnamefont {H.-S.}\ \bibnamefont {Lee}}, \
  and\ \bibinfo {author} {\bibfnamefont {S.}~\bibnamefont {Yun}},\ }\href
  {\doibase 10.1103/PhysRevLett.118.101802} {\bibfield  {journal} {\bibinfo
  {journal} {Phys. Rev. Lett.}\ }\textbf {\bibinfo {volume} {118}},\ \bibinfo
  {pages} {101802} (\bibinfo {year} {2017}{\natexlab{a}})},\ \Eprint
  {http://arxiv.org/abs/1611.01466} {arXiv:1611.01466 [hep-ph]} \BibitemShut
  {NoStop}%
\bibitem [{\citenamefont {Kaneta}\ \emph
  {et~al.}(2017{\natexlab{b}})\citenamefont {Kaneta}, \citenamefont {Lee},\
  and\ \citenamefont {Yun}}]{Kaneta:2017wfh}%
  \BibitemOpen
  \bibfield  {author} {\bibinfo {author} {\bibfnamefont {K.}~\bibnamefont
  {Kaneta}}, \bibinfo {author} {\bibfnamefont {H.-S.}\ \bibnamefont {Lee}}, \
  and\ \bibinfo {author} {\bibfnamefont {S.}~\bibnamefont {Yun}},\ }\href
  {\doibase 10.1103/PhysRevD.95.115032} {\bibfield  {journal} {\bibinfo
  {journal} {Phys. Rev. D}\ }\textbf {\bibinfo {volume} {95}},\ \bibinfo
  {pages} {115032} (\bibinfo {year} {2017}{\natexlab{b}})},\ \Eprint
  {http://arxiv.org/abs/1704.07542} {arXiv:1704.07542 [hep-ph]} \BibitemShut
  {NoStop}%
\bibitem [{\citenamefont {Hook}\ \emph {et~al.}(2021)\citenamefont {Hook},
  \citenamefont {Marques-Tavares},\ and\ \citenamefont
  {Ristow}}]{Hook:2021ous}%
  \BibitemOpen
  \bibfield  {author} {\bibinfo {author} {\bibfnamefont {A.}~\bibnamefont
  {Hook}}, \bibinfo {author} {\bibfnamefont {G.}~\bibnamefont
  {Marques-Tavares}}, \ and\ \bibinfo {author} {\bibfnamefont {C.}~\bibnamefont
  {Ristow}},\ }\href {\doibase 10.1007/JHEP06(2021)167} {\bibfield  {journal}
  {\bibinfo  {journal} {JHEP}\ }\textbf {\bibinfo {volume} {06}},\ \bibinfo
  {pages} {167} (\bibinfo {year} {2021})},\ \Eprint
  {http://arxiv.org/abs/2105.06476} {arXiv:2105.06476 [hep-ph]} \BibitemShut
  {NoStop}%
\bibitem [{\citenamefont {deNiverville}\ \emph {et~al.}(2018)\citenamefont
  {deNiverville}, \citenamefont {Lee},\ and\ \citenamefont
  {Seo}}]{deNiverville:2018hrc}%
  \BibitemOpen
  \bibfield  {author} {\bibinfo {author} {\bibfnamefont {P.}~\bibnamefont
  {deNiverville}}, \bibinfo {author} {\bibfnamefont {H.-S.}\ \bibnamefont
  {Lee}}, \ and\ \bibinfo {author} {\bibfnamefont {M.-S.}\ \bibnamefont
  {Seo}},\ }\href {\doibase 10.1103/PhysRevD.98.115011} {\bibfield  {journal}
  {\bibinfo  {journal} {Phys. Rev. D}\ }\textbf {\bibinfo {volume} {98}},\
  \bibinfo {pages} {115011} (\bibinfo {year} {2018})},\ \Eprint
  {http://arxiv.org/abs/1806.00757} {arXiv:1806.00757 [hep-ph]} \BibitemShut
  {NoStop}%
\bibitem [{\citenamefont {Choi}\ \emph {et~al.}(2020)\citenamefont {Choi},
  \citenamefont {Lee}, \citenamefont {Seong},\ and\ \citenamefont
  {Yun}}]{Choi:2018mvk}%
  \BibitemOpen
  \bibfield  {author} {\bibinfo {author} {\bibfnamefont {K.}~\bibnamefont
  {Choi}}, \bibinfo {author} {\bibfnamefont {S.}~\bibnamefont {Lee}}, \bibinfo
  {author} {\bibfnamefont {H.}~\bibnamefont {Seong}}, \ and\ \bibinfo {author}
  {\bibfnamefont {S.}~\bibnamefont {Yun}},\ }\href {\doibase
  10.1103/PhysRevD.101.043007} {\bibfield  {journal} {\bibinfo  {journal}
  {Phys. Rev. D}\ }\textbf {\bibinfo {volume} {101}},\ \bibinfo {pages}
  {043007} (\bibinfo {year} {2020})},\ \Eprint
  {http://arxiv.org/abs/1806.09508} {arXiv:1806.09508 [hep-ph]} \BibitemShut
  {NoStop}%
\bibitem [{\citenamefont {Biswas}\ \emph {et~al.}(2019)\citenamefont {Biswas},
  \citenamefont {Chatterjee}, \citenamefont {Gabrielli},\ and\ \citenamefont
  {Mele}}]{Biswas:2019lcp}%
  \BibitemOpen
  \bibfield  {author} {\bibinfo {author} {\bibfnamefont {S.}~\bibnamefont
  {Biswas}}, \bibinfo {author} {\bibfnamefont {A.}~\bibnamefont {Chatterjee}},
  \bibinfo {author} {\bibfnamefont {E.}~\bibnamefont {Gabrielli}}, \ and\
  \bibinfo {author} {\bibfnamefont {B.}~\bibnamefont {Mele}},\ }\href {\doibase
  10.1103/PhysRevD.100.115040} {\bibfield  {journal} {\bibinfo  {journal}
  {Phys. Rev. D}\ }\textbf {\bibinfo {volume} {100}},\ \bibinfo {pages}
  {115040} (\bibinfo {year} {2019})},\ \Eprint
  {http://arxiv.org/abs/1906.10608} {arXiv:1906.10608 [hep-ph]} \BibitemShut
  {NoStop}%
\bibitem [{\citenamefont {Kalashev}\ \emph {et~al.}(2019)\citenamefont
  {Kalashev}, \citenamefont {Kusenko},\ and\ \citenamefont
  {Vitagliano}}]{Kalashev:2018bra}%
  \BibitemOpen
  \bibfield  {author} {\bibinfo {author} {\bibfnamefont {O.~E.}\ \bibnamefont
  {Kalashev}}, \bibinfo {author} {\bibfnamefont {A.}~\bibnamefont {Kusenko}}, \
  and\ \bibinfo {author} {\bibfnamefont {E.}~\bibnamefont {Vitagliano}},\
  }\href {\doibase 10.1103/PhysRevD.99.023002} {\bibfield  {journal} {\bibinfo
  {journal} {Phys. Rev. D}\ }\textbf {\bibinfo {volume} {99}},\ \bibinfo
  {pages} {023002} (\bibinfo {year} {2019})},\ \Eprint
  {http://arxiv.org/abs/1808.05613} {arXiv:1808.05613 [hep-ph]} \BibitemShut
  {NoStop}%
\bibitem [{\citenamefont {Arias}\ \emph {et~al.}(2021)\citenamefont {Arias},
  \citenamefont {Arza}, \citenamefont {Jaeckel},\ and\ \citenamefont
  {Vargas-Arancibia}}]{Arias:2020tzl}%
  \BibitemOpen
  \bibfield  {author} {\bibinfo {author} {\bibfnamefont {P.}~\bibnamefont
  {Arias}}, \bibinfo {author} {\bibfnamefont {A.}~\bibnamefont {Arza}},
  \bibinfo {author} {\bibfnamefont {J.}~\bibnamefont {Jaeckel}}, \ and\
  \bibinfo {author} {\bibfnamefont {D.}~\bibnamefont {Vargas-Arancibia}},\
  }\href {\doibase 10.1088/1475-7516/2021/05/070} {\bibfield  {journal}
  {\bibinfo  {journal} {JCAP}\ }\textbf {\bibinfo {volume} {05}},\ \bibinfo
  {pages} {070} (\bibinfo {year} {2021})},\ \Eprint
  {http://arxiv.org/abs/2007.12585} {arXiv:2007.12585 [hep-ph]} \BibitemShut
  {NoStop}%
\bibitem [{\citenamefont {Helm}(1956)}]{Helm:1956zz}%
  \BibitemOpen
  \bibfield  {author} {\bibinfo {author} {\bibfnamefont {R.~H.}\ \bibnamefont
  {Helm}},\ }\href {\doibase 10.1103/PhysRev.104.1466} {\bibfield  {journal}
  {\bibinfo  {journal} {Phys. Rev.}\ }\textbf {\bibinfo {volume} {104}},\
  \bibinfo {pages} {1466} (\bibinfo {year} {1956})}\BibitemShut {NoStop}%
\bibitem [{\citenamefont {Aguilar-Arevalo}\ \emph
  {et~al.}(2009{\natexlab{b}})\citenamefont {Aguilar-Arevalo} \emph
  {et~al.}}]{MiniBooNE:2008hfu}%
  \BibitemOpen
  \bibfield  {author} {\bibinfo {author} {\bibfnamefont {A.~A.}\ \bibnamefont
  {Aguilar-Arevalo}} \emph {et~al.} (\bibinfo {collaboration} {MiniBooNE}),\
  }\href {\doibase 10.1103/PhysRevD.79.072002} {\bibfield  {journal} {\bibinfo
  {journal} {Phys. Rev. D}\ }\textbf {\bibinfo {volume} {79}},\ \bibinfo
  {pages} {072002} (\bibinfo {year} {2009}{\natexlab{b}})},\ \Eprint
  {http://arxiv.org/abs/0806.1449} {arXiv:0806.1449 [hep-ex]} \BibitemShut
  {NoStop}%
\bibitem [{\citenamefont {Bonesini}\ \emph {et~al.}(2001)\citenamefont
  {Bonesini}, \citenamefont {Marchionni}, \citenamefont {Pietropaolo},\ and\
  \citenamefont {Tabarelli~de Fatis}}]{Bonesini:2001iz}%
  \BibitemOpen
  \bibfield  {author} {\bibinfo {author} {\bibfnamefont {M.}~\bibnamefont
  {Bonesini}}, \bibinfo {author} {\bibfnamefont {A.}~\bibnamefont
  {Marchionni}}, \bibinfo {author} {\bibfnamefont {F.}~\bibnamefont
  {Pietropaolo}}, \ and\ \bibinfo {author} {\bibfnamefont {T.}~\bibnamefont
  {Tabarelli~de Fatis}},\ }\href {\doibase 10.1007/s100520100656} {\bibfield
  {journal} {\bibinfo  {journal} {Eur. Phys. J. C}\ }\textbf {\bibinfo {volume}
  {20}},\ \bibinfo {pages} {13} (\bibinfo {year} {2001})},\ \Eprint
  {http://arxiv.org/abs/hep-ph/0101163} {arXiv:hep-ph/0101163} \BibitemShut
  {NoStop}%
\bibitem [{\citenamefont {Schmitz}(2008)}]{Schmitz:2008zz}%
  \BibitemOpen
  \bibfield  {author} {\bibinfo {author} {\bibfnamefont {D.~W.}\ \bibnamefont
  {Schmitz}},\ }\emph {\bibinfo {title} {{A Measurement of Hadron Production
  Cross Sections for the Simulation of Accelerator Neutrino Beams and a Search
  for $\nu_\mu$ to $\nu_e$ Sscillations in the $\delta m^{2}$ about equals
  $1-eV^{2}$ Region}}},\ \href {\doibase 10.2172/935240} {Ph.D. thesis},\
  \bibinfo  {school} {Columbia U.} (\bibinfo {year} {2008})\BibitemShut
  {NoStop}%
\bibitem [{\citenamefont {Alwall}\ \emph {et~al.}(2014)\citenamefont {Alwall},
  \citenamefont {Frederix}, \citenamefont {Frixione}, \citenamefont {Hirschi},
  \citenamefont {Maltoni}, \citenamefont {Mattelaer}, \citenamefont {Shao},
  \citenamefont {Stelzer}, \citenamefont {Torrielli},\ and\ \citenamefont
  {Zaro}}]{Alwall:2014hca}%
  \BibitemOpen
  \bibfield  {author} {\bibinfo {author} {\bibfnamefont {J.}~\bibnamefont
  {Alwall}}, \bibinfo {author} {\bibfnamefont {R.}~\bibnamefont {Frederix}},
  \bibinfo {author} {\bibfnamefont {S.}~\bibnamefont {Frixione}}, \bibinfo
  {author} {\bibfnamefont {V.}~\bibnamefont {Hirschi}}, \bibinfo {author}
  {\bibfnamefont {F.}~\bibnamefont {Maltoni}}, \bibinfo {author} {\bibfnamefont
  {O.}~\bibnamefont {Mattelaer}}, \bibinfo {author} {\bibfnamefont {H.~S.}\
  \bibnamefont {Shao}}, \bibinfo {author} {\bibfnamefont {T.}~\bibnamefont
  {Stelzer}}, \bibinfo {author} {\bibfnamefont {P.}~\bibnamefont {Torrielli}},
  \ and\ \bibinfo {author} {\bibfnamefont {M.}~\bibnamefont {Zaro}},\ }\href
  {\doibase 10.1007/JHEP07(2014)079} {\bibfield  {journal} {\bibinfo  {journal}
  {JHEP}\ }\textbf {\bibinfo {volume} {07}},\ \bibinfo {pages} {079} (\bibinfo
  {year} {2014})},\ \Eprint {http://arxiv.org/abs/1405.0301} {arXiv:1405.0301
  [hep-ph]} \BibitemShut {NoStop}%
\bibitem [{\citenamefont {Patterson}\ \emph {et~al.}(2009)\citenamefont
  {Patterson}, \citenamefont {Laird}, \citenamefont {Liu}, \citenamefont
  {Meyers}, \citenamefont {Stancu},\ and\ \citenamefont
  {Tanaka}}]{Patterson:2009ki}%
  \BibitemOpen
  \bibfield  {author} {\bibinfo {author} {\bibfnamefont {R.~B.}\ \bibnamefont
  {Patterson}}, \bibinfo {author} {\bibfnamefont {E.~M.}\ \bibnamefont
  {Laird}}, \bibinfo {author} {\bibfnamefont {Y.}~\bibnamefont {Liu}}, \bibinfo
  {author} {\bibfnamefont {P.~D.}\ \bibnamefont {Meyers}}, \bibinfo {author}
  {\bibfnamefont {I.}~\bibnamefont {Stancu}}, \ and\ \bibinfo {author}
  {\bibfnamefont {H.~A.}\ \bibnamefont {Tanaka}},\ }\href {\doibase
  10.1016/j.nima.2009.06.064} {\bibfield  {journal} {\bibinfo  {journal} {Nucl.
  Instrum. Meth. A}\ }\textbf {\bibinfo {volume} {608}},\ \bibinfo {pages}
  {206} (\bibinfo {year} {2009})},\ \Eprint {http://arxiv.org/abs/0902.2222}
  {arXiv:0902.2222 [hep-ex]} \BibitemShut {NoStop}%
\bibitem [{\citenamefont {Wang}\ \emph {et~al.}(2015)\citenamefont {Wang},
  \citenamefont {Alvarez-Ruso},\ and\ \citenamefont {Nieves}}]{Wang:2014nat}%
  \BibitemOpen
  \bibfield  {author} {\bibinfo {author} {\bibfnamefont {E.}~\bibnamefont
  {Wang}}, \bibinfo {author} {\bibfnamefont {L.}~\bibnamefont {Alvarez-Ruso}},
  \ and\ \bibinfo {author} {\bibfnamefont {J.}~\bibnamefont {Nieves}},\ }\href
  {\doibase 10.1016/j.physletb.2014.11.025} {\bibfield  {journal} {\bibinfo
  {journal} {Phys. Lett. B}\ }\textbf {\bibinfo {volume} {740}},\ \bibinfo
  {pages} {16} (\bibinfo {year} {2015})},\ \Eprint
  {http://arxiv.org/abs/1407.6060} {arXiv:1407.6060 [hep-ph]} \BibitemShut
  {NoStop}%
\bibitem [{\citenamefont {Cortina~Gil}\ \emph {et~al.}(2021)\citenamefont
  {Cortina~Gil} \emph {et~al.}}]{NA62:2021bji}%
  \BibitemOpen
  \bibfield  {author} {\bibinfo {author} {\bibfnamefont {E.}~\bibnamefont
  {Cortina~Gil}} \emph {et~al.} (\bibinfo {collaboration} {NA62}),\ }\href
  {\doibase 10.1016/j.physletb.2021.136259} {\bibfield  {journal} {\bibinfo
  {journal} {Phys. Lett. B}\ }\textbf {\bibinfo {volume} {816}},\ \bibinfo
  {pages} {136259} (\bibinfo {year} {2021})},\ \Eprint
  {http://arxiv.org/abs/2101.12304} {arXiv:2101.12304 [hep-ex]} \BibitemShut
  {NoStop}%
\bibitem [{\citenamefont {Aguilar-Arevalo}\ \emph
  {et~al.}(2021{\natexlab{b}})\citenamefont {Aguilar-Arevalo} \emph
  {et~al.}}]{PIENU:2021clt}%
  \BibitemOpen
  \bibfield  {author} {\bibinfo {author} {\bibfnamefont {A.}~\bibnamefont
  {Aguilar-Arevalo}} \emph {et~al.} (\bibinfo {collaboration} {PIENU}),\ }\href
  {\doibase 10.1103/PhysRevD.103.052006} {\bibfield  {journal} {\bibinfo
  {journal} {Phys. Rev. D}\ }\textbf {\bibinfo {volume} {103}},\ \bibinfo
  {pages} {052006} (\bibinfo {year} {2021}{\natexlab{b}})},\ \Eprint
  {http://arxiv.org/abs/2101.07381} {arXiv:2101.07381 [hep-ex]} \BibitemShut
  {NoStop}%
\bibitem [{\citenamefont {Banerjee}\ \emph {et~al.}(2019)\citenamefont
  {Banerjee} \emph {et~al.}}]{Banerjee:2019pds}%
  \BibitemOpen
  \bibfield  {author} {\bibinfo {author} {\bibfnamefont {D.}~\bibnamefont
  {Banerjee}} \emph {et~al.},\ }\href {\doibase 10.1103/PhysRevLett.123.121801}
  {\bibfield  {journal} {\bibinfo  {journal} {Phys. Rev. Lett.}\ }\textbf
  {\bibinfo {volume} {123}},\ \bibinfo {pages} {121801} (\bibinfo {year}
  {2019})},\ \Eprint {http://arxiv.org/abs/1906.00176} {arXiv:1906.00176
  [hep-ex]} \BibitemShut {NoStop}%
\bibitem [{\citenamefont {Riordan}\ \emph {et~al.}(1987)\citenamefont {Riordan}
  \emph {et~al.}}]{Riordan:1987aw}%
  \BibitemOpen
  \bibfield  {author} {\bibinfo {author} {\bibfnamefont {E.~M.}\ \bibnamefont
  {Riordan}} \emph {et~al.},\ }\href {\doibase 10.1103/PhysRevLett.59.755}
  {\bibfield  {journal} {\bibinfo  {journal} {Phys. Rev. Lett.}\ }\textbf
  {\bibinfo {volume} {59}},\ \bibinfo {pages} {755} (\bibinfo {year}
  {1987})}\BibitemShut {NoStop}%
\bibitem [{\citenamefont {Lees}\ \emph {et~al.}(2014)\citenamefont {Lees} \emph
  {et~al.}}]{BaBar:2014zli}%
  \BibitemOpen
  \bibfield  {author} {\bibinfo {author} {\bibfnamefont {J.~P.}\ \bibnamefont
  {Lees}} \emph {et~al.} (\bibinfo {collaboration} {BaBar}),\ }\href {\doibase
  10.1103/PhysRevLett.113.201801} {\bibfield  {journal} {\bibinfo  {journal}
  {Phys. Rev. Lett.}\ }\textbf {\bibinfo {volume} {113}},\ \bibinfo {pages}
  {201801} (\bibinfo {year} {2014})},\ \Eprint {http://arxiv.org/abs/1406.2980}
  {arXiv:1406.2980 [hep-ex]} \BibitemShut {NoStop}%
\bibitem [{\citenamefont {for COHERENT~Collaboration}(2021)}]{coherentrecent}%
  \BibitemOpen
  \bibfield  {author} {\bibinfo {author} {\bibfnamefont {D.~P.}\ \bibnamefont
  {for COHERENT~Collaboration}},\ }\href
  {https://indico.phy.ornl.gov/event/126/} {\enquote {\bibinfo {title} {Probing
  cosmological dark matter with first bounds from coherent},}\ } (\bibinfo
  {year} {2021})\BibitemShut {NoStop}%
\bibitem [{\citenamefont {Abratenko}\ \emph
  {et~al.}(2021{\natexlab{b}})\citenamefont {Abratenko} \emph
  {et~al.}}]{MicroBooNE:2021gfj}%
  \BibitemOpen
  \bibfield  {author} {\bibinfo {author} {\bibfnamefont {P.}~\bibnamefont
  {Abratenko}} \emph {et~al.} (\bibinfo {collaboration} {MicroBooNE}),\ }\href
  {\doibase 10.1103/PhysRevD.104.052002} {\bibfield  {journal} {\bibinfo
  {journal} {Phys. Rev. D}\ }\textbf {\bibinfo {volume} {104}},\ \bibinfo
  {pages} {052002} (\bibinfo {year} {2021}{\natexlab{b}})},\ \Eprint
  {http://arxiv.org/abs/2101.04228} {arXiv:2101.04228 [hep-ex]} \BibitemShut
  {NoStop}%
\bibitem [{\citenamefont {Abratenko}\ \emph
  {et~al.}(2021{\natexlab{c}})\citenamefont {Abratenko} \emph
  {et~al.}}]{MicroBooNE:2021rmx}%
  \BibitemOpen
  \bibfield  {author} {\bibinfo {author} {\bibfnamefont {P.}~\bibnamefont
  {Abratenko}} \emph {et~al.} (\bibinfo {collaboration} {MicroBooNE}),\
  }\href@noop {} {\  (\bibinfo {year} {2021}{\natexlab{c}})},\ \Eprint
  {http://arxiv.org/abs/2110.14054} {arXiv:2110.14054 [hep-ex]} \BibitemShut
  {NoStop}%
\bibitem [{\citenamefont {Aguilar-Arevalo}\ \emph {et~al.}(2001)\citenamefont
  {Aguilar-Arevalo} \emph {et~al.}}]{LSND:2001aii}%
  \BibitemOpen
  \bibfield  {author} {\bibinfo {author} {\bibfnamefont {A.}~\bibnamefont
  {Aguilar-Arevalo}} \emph {et~al.} (\bibinfo {collaboration} {LSND}),\ }\href
  {\doibase 10.1103/PhysRevD.64.112007} {\bibfield  {journal} {\bibinfo
  {journal} {Phys. Rev. D}\ }\textbf {\bibinfo {volume} {64}},\ \bibinfo
  {pages} {112007} (\bibinfo {year} {2001})},\ \Eprint
  {http://arxiv.org/abs/hep-ex/0104049} {arXiv:hep-ex/0104049} \BibitemShut
  {NoStop}%
\bibitem [{us()}]{us}%
  \BibitemOpen
  \href@noop {} {}\Eprint {http://arxiv.org/abs/To appear, 2022} {To appear,
  2022} \BibitemShut {NoStop}%
\bibitem [{\citenamefont {Altmannshofer}\ \emph {et~al.}(2020)\citenamefont
  {Altmannshofer}, \citenamefont {Gori},\ and\ \citenamefont
  {Robinson}}]{Altmannshofer:2019yji}%
  \BibitemOpen
  \bibfield  {author} {\bibinfo {author} {\bibfnamefont {W.}~\bibnamefont
  {Altmannshofer}}, \bibinfo {author} {\bibfnamefont {S.}~\bibnamefont {Gori}},
  \ and\ \bibinfo {author} {\bibfnamefont {D.~J.}\ \bibnamefont {Robinson}},\
  }\href {\doibase 10.1103/PhysRevD.101.075002} {\bibfield  {journal} {\bibinfo
   {journal} {Phys. Rev. D}\ }\textbf {\bibinfo {volume} {101}},\ \bibinfo
  {pages} {075002} (\bibinfo {year} {2020})},\ \Eprint
  {http://arxiv.org/abs/1909.00005} {arXiv:1909.00005 [hep-ph]} \BibitemShut
  {NoStop}%
\bibitem [{\citenamefont {Borie}(2012)}]{Borie:2012tu}%
  \BibitemOpen
  \bibfield  {author} {\bibinfo {author} {\bibfnamefont {E.}~\bibnamefont
  {Borie}},\ }\href@noop {} {\  (\bibinfo {year} {2012})},\ \Eprint
  {http://arxiv.org/abs/1207.6651} {arXiv:1207.6651 [physics.gen-ph]}
  \BibitemShut {NoStop}%
\bibitem [{\citenamefont {Qattan}\ \emph {et~al.}(2005)\citenamefont {Qattan}
  \emph {et~al.}}]{Qattan:2004ht}%
  \BibitemOpen
  \bibfield  {author} {\bibinfo {author} {\bibfnamefont {I.~A.}\ \bibnamefont
  {Qattan}} \emph {et~al.},\ }\href {\doibase 10.1103/PhysRevLett.94.142301}
  {\bibfield  {journal} {\bibinfo  {journal} {Phys. Rev. Lett.}\ }\textbf
  {\bibinfo {volume} {94}},\ \bibinfo {pages} {142301} (\bibinfo {year}
  {2005})},\ \Eprint {http://arxiv.org/abs/nucl-ex/0410010}
  {arXiv:nucl-ex/0410010} \BibitemShut {NoStop}%
\bibitem [{\citenamefont {Holst}\ \emph {et~al.}(2022)\citenamefont {Holst},
  \citenamefont {Hooper},\ and\ \citenamefont {Krnjaic}}]{Holst:2021lzm}%
  \BibitemOpen
  \bibfield  {author} {\bibinfo {author} {\bibfnamefont {I.}~\bibnamefont
  {Holst}}, \bibinfo {author} {\bibfnamefont {D.}~\bibnamefont {Hooper}}, \
  and\ \bibinfo {author} {\bibfnamefont {G.}~\bibnamefont {Krnjaic}},\ }\href
  {\doibase 10.1103/PhysRevLett.128.141802} {\bibfield  {journal} {\bibinfo
  {journal} {Phys. Rev. Lett.}\ }\textbf {\bibinfo {volume} {128}},\ \bibinfo
  {pages} {141802} (\bibinfo {year} {2022})},\ \Eprint
  {http://arxiv.org/abs/2107.09067} {arXiv:2107.09067 [hep-ph]} \BibitemShut
  {NoStop}%
\bibitem [{\citenamefont {de~Swart}\ \emph {et~al.}(1997)\citenamefont
  {de~Swart}, \citenamefont {Rentmeester},\ and\ \citenamefont
  {Timmermans}}]{deSwart:1997ep}%
  \BibitemOpen
  \bibfield  {author} {\bibinfo {author} {\bibfnamefont {J.~J.}\ \bibnamefont
  {de~Swart}}, \bibinfo {author} {\bibfnamefont {M.~C.~M.}\ \bibnamefont
  {Rentmeester}}, \ and\ \bibinfo {author} {\bibfnamefont {R.~G.~E.}\
  \bibnamefont {Timmermans}},\ }\href@noop {} {\bibfield  {journal} {\bibinfo
  {journal} {PiN Newslett.}\ }\textbf {\bibinfo {volume} {13}},\ \bibinfo
  {pages} {96} (\bibinfo {year} {1997})},\ \Eprint
  {http://arxiv.org/abs/nucl-th/9802084} {arXiv:nucl-th/9802084} \BibitemShut
  {NoStop}%
\bibitem [{\citenamefont {Dutta}\ \emph {et~al.}(2021)\citenamefont {Dutta},
  \citenamefont {Ghosh}, \citenamefont {Huang},\ and\ \citenamefont
  {Kumar}}]{Dutta:2021afo}%
  \BibitemOpen
  \bibfield  {author} {\bibinfo {author} {\bibfnamefont {B.}~\bibnamefont
  {Dutta}}, \bibinfo {author} {\bibfnamefont {S.}~\bibnamefont {Ghosh}},
  \bibinfo {author} {\bibfnamefont {P.}~\bibnamefont {Huang}}, \ and\ \bibinfo
  {author} {\bibfnamefont {J.}~\bibnamefont {Kumar}},\ }\href@noop {} {\
  (\bibinfo {year} {2021})},\ \Eprint {http://arxiv.org/abs/2105.07655}
  {arXiv:2105.07655 [hep-ph]} \BibitemShut {NoStop}%
\end{thebibliography}%

\end{document}